\documentclass[twocolumn]{aastex701}
\usepackage{enumitem}

\begin{document}

\title{Delayed radio emission in tidal disruption events from collisions of outflows driven by disk instabilities}
\author[0000-0003-2872-5153]{Samantha C. Wu}\thanks{Carnegie Theoretical Astrophysics Center Fellow}
\affiliation{The Observatories of the Carnegie Institution for Science, Pasadena, CA 91101, USA}
\affiliation{Center for Interdisciplinary Exploration \& Research in Astrophysics (CIERA), Physics \& Astronomy, Northwestern University, Evanston, IL 60202, USA}
\email[show]{swu@carnegiescience.edu}  

\author[0000-0002-6347-3089]{Daichi Tsuna}
\affiliation{TAPIR, Mailcode 350-17, California Institute of Technology, Pasadena, CA 91125, USA}
\affiliation{Research Center for the Early Universe (RESCEU), School of Science, The University of Tokyo,  Bunkyo-ku, Tokyo 113-0033, Japan}
\affiliation{Center for Astrophysics $|$ Harvard \& Smithsonian, 60 Garden St, Cambridge, MA 02138, USA}
\email{tsuna@caltech.edu}

\author[0000-0001-6350-8168]{Brenna Mockler}\thanks{Carnegie Theoretical Astrophysics Center Fellow}
\affiliation{The Observatories of the Carnegie Institution for Science, Pasadena, CA 91101, USA}
\affiliation{Department of Physics and Astronomy, University of California, Davis, CA 95616, USA}
\email{bmockler@carnegiescience.edu}

\author[0000-0001-6806-0673]{Anthony L. Piro}
\affiliation{The Observatories of the Carnegie Institution for Science, Pasadena, CA 91101, USA}
\email{piro@carnegiescience.edu}

\begin{abstract}
     Delayed radio emission has been associated with a growing proportion of tidal disruption events (TDEs). For many events, the radio synchrotron emission is inferred to originate from the interaction of mildly-relativistic outflows, launched with delay times of $\sim 100$--$1000$~d after the TDE optical peak. The mechanism behind these outflows remains uncertain, but may relate to instabilities or state transitions in the accretion disk formed from the TDE. We model the radio emission powered by the collision of mass outflows (``flares") from TDE accretion disks, considering scenarios in which two successive disk flares collide with each other, as well as collisions between the outflow and the circumnuclear medium (CNM).  For flare masses of $\sim 0.01$--$0.1\, M_{\odot}$, varied CNM densities, and different time intervals between ejected flares, we demonstrate that the shocks formed by the collisions have velocities $0.05c$ -- $0.3c$ at $\sim 10^{17}$ cm and power bright radio emission of $L_{\nu} \sim 10^{27}$--$10^{30}\,  \mathrm{erg}\, \mathrm{s}^{-1}\, \mathrm{Hz}^{-1}$, consistent with the properties inferred for observed events. We quantify how the typical peak timescale and flux varies for different properties of our models, and compare our model predictions to a selection of TDEs with delayed radio emission. Our models successfully reproduce the light curves and spectral energy distributions for several events, supporting the idea that delayed outflows from disk instabilities and state transitions can power late-time radio emission in TDEs. 
\end{abstract}

\section{Introduction}
When a star passing by a supermassive black hole (SMBH) approaches too closely and becomes torn apart by the tidal force of the SMBH, an energetic electromagnetic transient called a tidal disruption event (TDE) ensues \citep{hills1975,rees1988,evans1989,phinney1989}. The light from a TDE encodes information about the black hole mass and host environment, properties of the disrupted star, and the physics of accretion \citep{guillochon2013,mockler2019,bonnerot2021}. Emission from TDEs has been detected across optical, UV, X-ray, and radio wavelengths \citep{auchettl2017,gezari2021,hammerstein2023,cendes2024,goodwin2025arXiv}. 

In particular, radio observations enrich the picture of TDEs by illuminating processes that expel outflows from the disruption, thereby providing insight into the hydrodynamical processes taking place in the SMBH environment after a TDE \citep{alexander2020}. 
For example, jetted TDEs ($\nu L_\nu\gtrsim 10^{40}$ erg s$^{-1}$) generate powerful radio emission via a relativistic jet (e.g. Swift J1644+57, \citealt{zauderer2011,zauderer2013,cendes2021b} and AT2022cmc, \citealt{andreoni2022}).  
For radio detections in the first months after the optical discovery of the TDE, the mechanism powering this prompt emission is typically interpreted as the interaction between the surrounding medium and material that was ejected around the time of the TDE. 
Among non-jetted TDEs (with $\nu L_\nu < 10^{39}$ erg s$^{-1}$), prompt radio emission associated with a non-relativistic outflow has been detected in several events, including ASASSN-14li, AT2019qiz, ASASSN-19bt, AT2019dsg, and others \citep{vanvelzen2016,alexander2016,alexander2020,cendes2021,goodwin2023,hu2025,alexander25}.

In addition, recent studies have reported the discovery of radio emission that is more delayed relative to the time of optical discovery, which exhibit diverse properties and potentially stem from varied origins \citep{horesh2021,horesh2021b,cendes2022}. 
Many non-jetted TDEs appear to show such delayed radio emission \citep{alexander25}.
\cite{cendes2024} report radio emission detected years after discovery for $\sim 40$\% of optically selected TDEs, inferring evidence for mildly relativistic outflows launched with delays of $\sim$100s--1000s of days after the TDE. \cite{goodwin2025arXiv} identify radio emission in half of an X-ray selected sample of TDEs, coincident with the peak X-ray luminosity and possibly delayed from the optical detection of a TDE. The radio emission exhibited for certain events, such as AT2018hyz \citep{cendes2022}, may be associated with off-axis jets \citep{sfaradi2024,matsumoto2023}, or the delayed radio rebrightening could arise from misaligned jet precession \citep{teboul2023,lu2024}. However, \cite{cendes2024} disfavor the scenario of off-axis jets for the observed outflow velocities and radio luminosities of their sample. 

In the context of both prompt emission and delayed emission, non-jetted TDEs sometimes exhibit  multiple flares of radio emission \citep[e.g., AT2019qiz, ASASSN-15oi, AT2024tvd, ASASSN-19bt, and others;][]{cendes2021,goodwin2023, christy2024,hajela2025,sfaradi2025,alexander25}. From these events, there are indications that the second flare tends to carry more energy than the first \citep{hajela2025,goodwin2025}. Though it is as yet unclear what causes the flares, they indicate that separate incidents of energy injection can power repeated radio emission in these TDEs, possibly from different underlying mechanisms. 


Despite evidence that delayed radio emission due to mildly relativistic outflows from TDEs may commonly occur, the origins of these outflows remain uncertain. In the context of the TDE picture, mechanisms to launch fast outflows may relate to the formation and evolution of an accretion disk around the SMBH. For instance, optical/UV emission observed in the years after a TDE indicates the presence of a persistent disk \citep[e.g.,][]{vanvelzen19, mummery2021}. Ejecta from the stream-stream collision process that is thought to form the accretion disk could induce outflows from an early phase of the TDE, which may power radio emission on a variety of timescales and luminosities as the outflows interact with the surrounding medium \citep{lu2020,hajela2025}. At later times, mass ejections from a TDE accretion disk may be associated with accretion state changes \citep{wevers2021,sfaradi2022,hajela2025}, which in particular may cause disk instabilities that lead to outflows via super-Eddington accretion \citep{lu2022,piro2025}. 

In this work, we explore the properties of the emission generated by the collisions of such instability-driven disk outflows. We consider scenarios where two successive outflows collide with each other, as well as when the outflows collide with dense surrounding material. By modeling the hydrodynamical evolution of the shock formed by the collision, we predict the velocity of the radio-emitting shock. We also calculate the resultant light curves and spectral energy distributions (SEDs), evaluating the typical timescales and luminosities produced from the different types of outflow collision models. Comparison between our models and a selection of observed TDEs reveals that the collision of disk outflows is able to power luminous radio emission consistent with the delayed radio emission that has been observed for many TDEs.  In particular, the collision of multiple flares produces rapidly evolving light curves that may be relevant to events exhibiting multiple episodes of sharply rising and declining emission. 
    
\begin{table*}
\centering
\hspace{-2cm} 
\begin{tabular}{ccccccc}
\hline
Label & $M_{\rm fl}$ $(M_{\odot})$ & $v_{\rm min}$ (c) & $v_{\rm max}$ (c) & $\Delta t$ (yr) &  $n_0$ (cm$^{-3}$) & Model Type \\
\hline

$M0.01\,  \Delta t\, 0.1$ & 0.01 & 0.04 & 0.4 & 0.1 & N/A & Flare+Flare \\
$M0.01\,  \Delta t\, 0.3$ & 0.01 & 0.04 & 0.4 & 0.3 & N/A & Flare+Flare \\
$M0.01\,  \Delta t\, 1.0$ & 0.01 & 0.04 & 0.4 & 1.0 & N/A & Flare+Flare \\
$M0.01\,  \Delta t\, 2.0$ & 0.01 & 0.04 & 0.4 & 2.0 & N/A & Flare+Flare \\
$M0.1\,\Delta t\, 0.1$ & 0.1 & 0.04 & 0.4 & 0.1 & N/A & Flare+Flare \\
$M0.1\,  \Delta t\, 0.3$ & 0.1 & 0.04 & 0.4 & 0.3 & N/A & Flare+Flare \\
$M0.1\,  \Delta t\, 1.0$ & 0.1 & 0.04 & 0.4 & 1.0 & N/A & Flare+Flare \\
$M0.1\,  \Delta t\, 2.0$ & 0.1 & 0.04 & 0.4 & 2.0 & N/A & Flare+Flare \\
\hline 

$\alpha 0 M0.01\, n_0 10$ & 0.01 & 0.04 & 0.4 & N/A & 10 & Flare+$\alpha0$CNM \\
$\alpha 0 M0.01\, n_0 30$ & 0.01 & 0.04 & 0.4 & N/A & 30 & Flare+$\alpha0$CNM \\
$\alpha 0 M0.01\, n_0 10^2$ & 0.01 & 0.04 & 0.4 & N/A & $10^2$ & Flare+$\alpha0$CNM \\
$\alpha 0 M0.01\, n_0 150$ & 0.01 & 0.04 & 0.4 & N/A & $150$ & Flare+$\alpha0$CNM \\
$\alpha 0 M0.01\, n_0 10^3$ & 0.01 & 0.04 & 0.4 & N/A & $10^3$ & Flare+$\alpha0$CNM \\
$\alpha 0 M0.1\, n_0 10$ & 0.1 & 0.04 & 0.4 & N/A & 10 & Flare+$\alpha0$CNM \\
$\alpha 0 M0.1\, n_0 30$ & 0.1 & 0.04 & 0.4 & N/A & 30 & Flare+$\alpha0$CNM \\
$\alpha 0 M0.1\, n_0 10^2$ & 0.1 & 0.04 & 0.4 & N/A & $10^2$ & Flare+$\alpha0$CNM \\
$\alpha 0 M0.1\, n_0 10^3$ & 0.1 & 0.04 & 0.4 & N/A & $10^3$ & Flare+$\alpha0$CNM \\
\hline 
$\alpha 2.5 M0.01\, n_0 10$ & 0.01 & 0.04 & 0.4 & N/A & $10$ & Flare+$\alpha2.5$CNM \\
$\alpha 2.5 M0.01\, n_0 30$ & 0.01 & 0.04 & 0.4 & N/A & $30$ & Flare+$\alpha2.5$CNM \\
$\alpha 2.5 M0.01\, n_0 10^2$ & 0.01 & 0.04 & 0.4 & N/A & $10^2$ & Flare+$\alpha2.5$CNM \\
$\alpha 2.5 M0.01\, n_0 10^3$ & 0.01 & 0.04 & 0.4 & N/A & $10^3$ & Flare+$\alpha2.5$CNM \\
$\alpha 2.5 M0.1\, n_0 10$ & 0.1 & 0.04 & 0.4 & N/A & $10$ & Flare+$\alpha2.5$CNM \\
$\alpha 2.5 M0.1\, n_0 30$ & 0.1 & 0.04 & 0.4 & N/A & $30$ & Flare+$\alpha2.5$CNM \\
$\alpha 2.5 M0.1\, n_0 10^2$ & 0.1 & 0.04 & 0.4 & N/A & $10^2$ & Flare+$\alpha2.5$CNM \\
$\alpha 2.5 M0.1\, n_0 10^3$ & 0.1 & 0.04 & 0.4 & N/A & $10^3$ & Flare+$\alpha2.5$CNM \\
\hline \\[-1mm]
\end{tabular}
\caption{Properties of selected models presented throughout this work.}
\label{tab:allmodels}
\end{table*}

\section{Radio emission from flare collisions}
We examine the properties of radio emission powered by the collision of mass ejections from a TDE, which form outflows that we call flares. Here, we consider two main cases: first, that the collision occurs between two successive flares that are ejected separated by a time interval $\Delta t$; and second, that the collision occurs between the flare and the surrounding circumnuclear medium (CNM). When we explore the collision of two successive flares, we assume that the two flares have identical flare masses and velocity ranges, which is a good approximation to the behavior of the flares from \cite{piro2025} from $\sim 1$ yr onward (see their Figure 9).  Table \ref{tab:allmodels} lists the names and properties of each of the models.

Throughout, we refer to calculations from the collision of two successive outflows by the label Flare+Flare. 
For the collision of an outflow with the CNM, we consider CNM densities that vary with radius as $\rho_{\rm CNM} \propto r^{-\alpha}$. In this work, we explore $\alpha=0$ (constant density CNM) and $\alpha=2.5$, which spans the range of observationally inferred slopes for a variety of TDEs \citep[e.g.,][]{alexander2016,cendes2021,goodwin2022,burn2025arXiv}.
For the constant density case, the models are labeled as Flare+$\alpha0$CNM, and for the case of $\rho_{\rm CNM} \propto r^{-2.5}$, the models are labeled as Flare+$\alpha2.5$CNM.

\subsection{Outflow properties}

Delayed radio emission from TDEs has been posited to be powered by the interaction of fast-moving outflows with dense material \citep{cendes2024}. 
Such outflows may be launched during an early phase of the TDE, such as during the circularization process \citep{bonnerot2021b,huang2024}. Other models suggest that super-Eddington outflows may be ejected from a TDE accretion disk during state transitions, such as between super-Eddington and sub-Eddington disk states \citep{alexander25,piro2025}. Here, we focus on modeling the interactions of outflows launched due to thermal disk instabilities when the disk is in a high state of accretion. 

Accordingly, our choice of flare density profile is motivated by a scenario where a flare is launched from the accretion disk of a SMBH. The material leaves the disk with range of velocities $v_{\rm min} < v < v_{\rm max}$ set by the inner and outer radii of the disk; these velocities are proportional to the escape speed from the disk $v \propto \sqrt{G M_{\rm BH}/r_{\rm BH}}\propto r_{\rm BH}^{-1/2}$, where $r_{\rm BH}$ is the distance from the SMBH. As in \cite{piro2025}, we assume that the disk has a power-law inflow rate of $\dot{M}_{\rm in} \propto r_{\rm BH}^s \propto v^{-2s}$ \citep{blandford99}. We adopt $s = 0.5$, as motivated by recent MHD simulations of radiatively-inefficient accretion flows \citep{guo2024,cho2024}.

To derive the density profile of the flare, we consider the mass of a spherical shell ejected from the disk with velocity in the range of ($v$, $v+d v$) with a duration $t_{\rm fl}$:
\begin{equation}
    t_{\rm fl} \frac{d \dot{M}_{\mathrm{in}}}{dv} \, dv = 4\pi r^2 \rho_{\rm fl} \, dr 
\end{equation}
Once the flare reaches a phase of homologous expansion such that $v=r/t$, using the inflow rate above gives
\begin{eqnarray}
    v^{-2s-1} \, dv &\propto (vt)^2 \rho\, t dv \nonumber \\ 
 \rho_{\rm fl} (v,t) &\propto v^{-2s-3}t^{-3}.
\end{eqnarray}
By scaling to $v_{\rm max}$, we find the following flare density profile as a function of radius $r$ at a given time $t$ from the moment the outflow is launched \citep{piro2025}:
\begin{equation}
\label{eq:rhoflare}
    \rho_{\rm fl}(r,t) = A t^{-3}\left(\frac{r}{v_{\rm max} t} \right)^{-2s-3}
\end{equation}
for $v_{\rm min} < r/t < v_{\rm max}$.
We set $v_{\rm min}$ and $v_{\rm max}$ and choose the flare mass $M_{\rm fl}$ in order to specify the normalization constant $A$: 
\begin{equation}
    A = \frac{sM_{\rm fl}}{2\pi v_{\rm max}^3\left[ \left(\frac{v_{\rm max}}{v_{\rm min}}\right)^{2s} - 1 \right]}
\end{equation}
The kinetic energy of the flare that is ejected is thus
\begin{equation}
    E_{\rm fl} = \frac{\pi A v_{\rm max}^5}{1-s}\left[1-\left(\frac{v_{\rm min}}{v_{\rm max}}\right)^{2-2s}\right].
\end{equation}
For our fiducial parameters of $s=0.5$, $v_{\rm max}=0.4c$, $v_{\rm min}=0.04c$, the initial flare energy is $E_{\rm fl} = 1.4\times10^{50}\, \mathrm{erg}\, \left(M_{\rm fl}/0.01\, M_{\odot}\right)$.

The framework outlined in Sections \ref{sec:shockevol} and \ref{sec:synchrotronframework} is independent of this choice of flare density profile.

\subsection{Shock evolution}
\label{sec:shockevol}
In this section, we describe our methods for calculating the hydrodynamical evolution of the shock formed by the collision of mass ejections from a TDE. 
A shock will form once the flare collides with a flare that was launched earlier by a time interval $\Delta t$, or alternatively when the flare collides with the CNM. 
We assume the shocked region is a thin shell with radius $r_{\rm sh}$ and velocity $v_{\rm sh}$, and from there we calculate the propagation of the shock formed by the collision via mass and momentum conservation:
\begin{eqnarray}
    \frac{dM_{\rm sh}}{dt} &=& 4\pi r_{\rm sh}^2[\rho_{2}(v_{2}-v_{\rm sh})+\rho_{1}(v_{\rm sh}-v_{1})] \nonumber \\ \nonumber\\
    M_{\rm sh}\frac{dv_{\rm sh}}{dt} &=& 4\pi r_{\rm sh}^2 [\rho_{2}(v_{2}-v_{\rm sh})^2 - \rho_{1}(v_{\rm sh}-v_{1})^2] \nonumber \\ \nonumber\\
    \frac{dr_{\rm sh}}{dt} &=& v_{\rm sh}
    \label{eq:dvshdt}
\end{eqnarray}

For the Flare+Flare case, the shock is formed by a collision between two successive flares. We substitute $v_1$ with $v_{\rm fl, 1} = r_{\rm sh}/(t + \Delta t)$ and $\rho_1$ with $\rho_{\rm fl,1} = \rho_{\rm fl}(r_{\rm sh},t+\Delta t)$, since the first flare has had an extra time interval of $\Delta t$ to expand. For $v_2$ and $\rho_2$, we substitute $v_2 = v_{\rm fl,2} = r_{\rm sh}/t$ and $\rho_2 =  \rho_{\rm fl,2} = \rho_{\rm fl}(r_{\rm sh},t)$. In our calculations, we measure the time from the launch of the second flare. The collision will take place at a time $t_{0} = v_{\rm min} \Delta t/(v_{\rm max}-v_{\rm min})$, when the fastest part of the second flare catches up to the slowest part of the first flare. This also sets the initial condition for $r_{\rm sh}$: $r_{\rm sh,0} = v_{\rm max}t_{0}$.

For the Flare+$\alpha0$CNM and Flare+$\alpha2.5$CNM models, the shock is formed by the collision between a flare and the CNM. We substitute $v_1=0$ and $\rho_1=\rho_{\rm CNM}(r_{\rm sh})$ for the CNM. For the flare, we again use $v_2 = v_{\rm fl,2} = r_{\rm sh}/t$ and $\rho_2 = \rho_{\rm fl}(r_{\rm sh},t)$. The initial shock radius is $r_{\rm sh, 0} = v_{\rm max}t_0$, and we choose $t_0 = 10^{-3}$ yr.

In all cases, the initial condition for $M_{\rm sh}$ is $M_{\rm sh,0}=0$ at $t=t_0$. The initial shock velocity $v_{\rm sh,0}$ is set by the ram pressure balance of either the two flares or the flare and the CNM at $t_0$, i.e. where $\rho_1 (v_{\rm sh,0}-v_1)^2=\rho_2 (v_{2}-v_{\rm sh,0})^2$.

\subsection{Radio Emission}
\label{sec:synchrotronframework}
We outline the framework used to calculate the radio emission from the flare collisions below. The emission from the shock formed by the collision depends on both the density $\rho_i$ of the medium through which the shock passes, and the relative velocity $v_i$ of the shock compared to the velocity of the material upstream of the shock.  

For the Flare+Flare models, we perform this calculation for both the forward shock passing through the first flare and the reverse shock traversing the second flare: for the first flare, we calculate the emission using $\rho_i = \rho_{\rm fl,1}$ and $v_{i} = v_{\rm sh}-v_{\rm fl,1}$, and for the second flare, we use $\rho_i = \rho_{\rm fl,2}$ and $v_{i} = v_{\rm fl,2}-v_{\rm sh}$. For the Flare+$\alpha0$CNM and Flare+$\alpha2.5$CNM models, we assume that the forward shock interacting with the CNM dominates the radio emission, so we calculate the emission for only the forward shock using $\rho_i = \rho_{\rm CNM}$ and $v_i = v_{\rm sh}$.

The electrons in the shocked region cool by emitting synchrotron radiation. We take the magnetic field strength, parameterized by an efficiency parameter $\epsilon_B$ relative to the upstream ram pressure, to be
\begin{equation}
\label{eq:Bfield}
    B=\sqrt{8\pi \epsilon_B \rho_i v_{i}^2}.
\end{equation}

We assume that the number density of relativistic electrons injected into the shocked region can be described by a power-law energy distribution in Lorentz factor, $dn(\gamma_e)/d\gamma_e=n_{0,e} \gamma_e^{-p}$ ($p>2$), as expected for diffusive shock acceleration. The spectral index of the radio emission is correlated with the power-law index $p$; in this work, we have assumed $p=3$, which has been inferred for several TDEs \citep[e.g.,][]{cendes2024,goodwin2025arXiv}.
The normalization of the electron distribution $n_{0,e}$ is scaled by the parameter $\epsilon_e$, which describes the fraction of the energy density of relativistic electrons compared to the ram pressure: 
\begin{eqnarray}\label{eq:electron_distribution}
    \int_{\gamma_{\rm min}}^{\infty} (\gamma_e m_e c^2)\frac{dn}{d\gamma_e}d\gamma_e =\epsilon_e \rho_{i} v_{i}^2,
\end{eqnarray}
where $m_e$ is the electron mass. For $p=3$ this leads to a normalization of
\begin{eqnarray}
    n_{0,e} = \frac{(\epsilon_e\gamma_{\rm min})\rho_{i}v_{i}^2}{m_e c^2}.
    \label{eq:electron_n0}
\end{eqnarray} 
We assume the non-thermal electron spectrum extends to $\gamma_{\rm min}=1$ \citep{Chevalier98}, and adopt $\epsilon_B=\epsilon_e=0.1$  to calculate the light curves in this work. 

\begin{figure*}
    \centering
    \includegraphics[width=\textwidth]{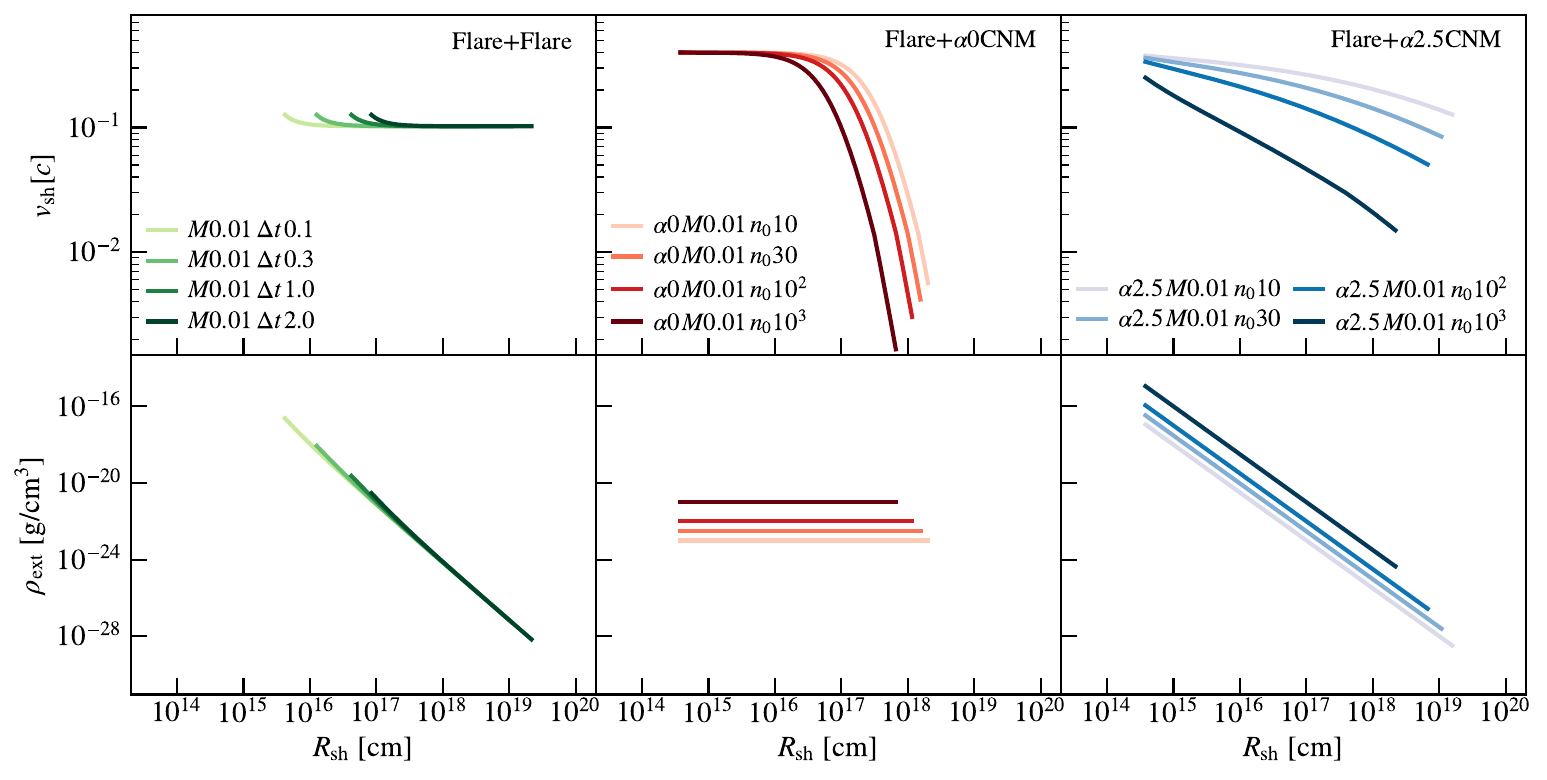}
    \caption{Evolution of the shock velocity and density ahead of the forward shock as a function of shock radius, for a set of flares with flare mass $M_{\rm fl} = 0.01\, M_{\odot}$ and flare velocity range of $v_{\rm min} = 0.04 c$, $v_{\rm max} = 0.4 c$. The time interval $\Delta t$ between two flares is varied in the first column (Flare+Flare), and in the last two columns, the normalization of the circumnuclear medium (CNM) density $n_0$ is varied. }
    \label{fig:M0p01_shellevol}
\end{figure*}

To calculate the synchrotron emission, we follow the methods of \cite{wu2025}, Equations 16--25. For the free-free absorption (FFA), we assume a fully ionized gas with hydrogen mass fraction $X=0.7$, helium mass fraction $Y=0.3$, and electron temperature $T_{\rm e}\sim 10^4$ K. 
In the models of the collision of two flares, we integrate the FFA out to the outer edge of the flare that was ejected first, $r_{\rm out}(t) = v_{\rm max}(t+\Delta t)$. For the models of a flare colliding with CNM, a reasonable value for $r_{\rm out}$ is set at the radius of the black hole's sphere of influence, $r_{\rm out} \sim 4\, \mathrm{pc}\, (M_{\rm BH}/10^7 M_{\odot})^{0.5}$ \citep{greene2020}. We assume $r_{\rm out}=4$ pc in our models, although we find that the FFA in the Flare+$\alpha2.5$CNM models is negligible from very early times onward and does not affect the peaks in our light curves. 
For the Flare+$\alpha0$CNM models, FFA is not important at the CNM densities we explore in this work. 

We obtain the observed spectrum as
\begin{eqnarray}
    L_{\nu, \rm obs} \approx L_{\nu, \rm syn}\exp(-\tau_{\rm ff}) \frac{1-\exp(-{\tau_{\rm ssa}})}{\tau_{\rm ssa}},
\end{eqnarray}
where $\tau_{\rm ssa}$ is the synchrotron self-absorption (SSA) optical depth, $\tau_{\rm ff}$ is the FFA optical depth, and $L_{\nu, \rm syn}$ is the unabsorbed synchrotron luminosity (Equations 20--24 in \citealt{wu2025}). 

\begin{figure*}
    \centering
    \includegraphics[width=\textwidth]{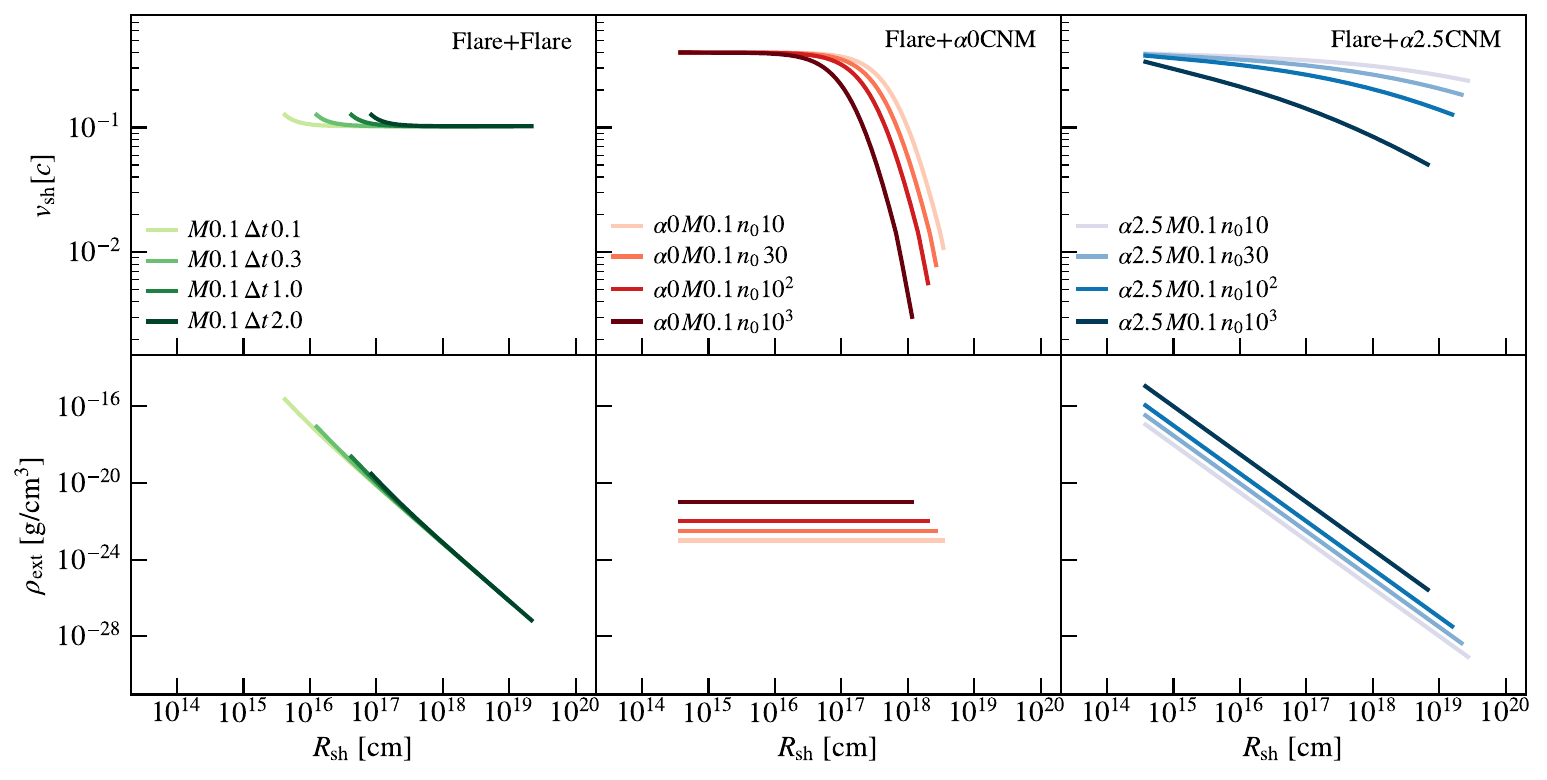}
    \caption{Same as Figure \ref{fig:M0p01_shellevol}, for flare mass $M_{\rm fl} = 0.1\, M_{\odot}$.}
    \label{fig:M0p1_shellevol}
\end{figure*}

\section{Results}

\subsection{Hydrodynamical evolution}

Figures \ref{fig:M0p01_shellevol} and \ref{fig:M0p1_shellevol} show the evolution of the density ahead of the shocked shell formed by the collision and the velocity of the shocked shell, as a function of the shock radius.
For the models shown in Figures \ref{fig:M0p01_shellevol} and \ref{fig:M0p1_shellevol}, the velocity range of $(v_{\rm min}, v_{\rm max})=(0.04c,0.4c)$ is motivated by the disk model of \cite{piro2025}.
This choice leads to shock velocities that evolve through the range inferred from late-time radio emission in TDEs \citep{cendes2024,goodwin2025arXiv}. We choose to explore two fiducial flare masses of $M_{\rm fl}=0.01\, M_{\odot}$ and $M_{\rm fl}=0.1\, M_{\odot}$ in each of these figures respectively.

\subsubsection{Collision of two flares}


In the case of two colliding flares (the Flare+Flare column in each of Figures \ref{fig:M0p01_shellevol} and \ref{fig:M0p1_shellevol}), we vary the time between the launching of each outflow, $\Delta t$. This leads to different velocity evolution as a function of radius. The time intervals of $0.1$--$2$ yr are motivated by the rough timescales between successive disk instabilities found by \cite{piro2025}.
Although we calculate both the forward and reverse shocks for the Flare+Flare case, we show only the forward shock properties in Figures \ref{fig:M0p01_shellevol} and \ref{fig:M0p1_shellevol}. As a result, the density ahead of the forward shock $\rho_{\rm ext}$ is set by the density profile of the first flare, which decreases with radius (Equation \ref{eq:rhoflare}).

The time and radius of the collision depend linearly on $\Delta t$:
\begin{equation}
    t_{\rm coll} = \frac{v_{\rm min} \Delta t}{v_{\rm max}-v_{\rm min}},~ r_{\rm coll}=\frac{v_{\rm max} v_{\rm min} \Delta t}{v_{\rm max}-v_{\rm min}}
\end{equation}
This sets the initial radius and time of the calculations shown in the Flare+Flare column for Figures \ref{fig:M0p01_shellevol} and \ref{fig:M0p1_shellevol}. 
Furthermore, a smaller $\Delta t$ corresponds to a larger upstream density ahead of the shell, as the first flare has had less time to expand to larger radii.
More massive flares (e.g. Figure \ref{fig:M0p1_shellevol}) also lead to systematically larger densities ahead of the shock. 
The velocity of the shocked region remains around $\sim 0.1c$ for all the Flare+Flare models out to large radii; this is consistent with the inferred outflow speeds from, e.g., \cite{cendes2024} and \cite{goodwin2025}, which are derived from equipartition arguments and likely lower limits.


\subsubsection{Collision of a flare with the CNM}

In the case of a flare colliding with the CNM, we vary the normalization of the number density of the surrounding CNM, $n_0$. We scale the density of the CNM to this normalization by $\rho_{\rm CNM} = m_p n_0 (r/r_0)^{-\alpha}$, where we choose $r_0=10^{17}$ cm. 

In this work, we explore two values of $\alpha$: for the constant density case, $\alpha=0$ (Flare+$\alpha0$CNM models), and $\alpha=2.5$ for a power-law decline in density (Flare+$\alpha2.5$CNM models). Our choice of this power law index is motivated by previous inferences of the circumnuclear density around TDEs \citep[e.g.][]{alexander2016,cendes2021}. 

As we normalize the CNM density at a radius of $r_0=10^{17}$ cm, the values of $n_0$ reported in Table \ref{tab:allmodels} for the Flare+$\alpha2.5$CNM case therefore refer to the density of the CNM at that radius. As a result, the initial density encountered by the flare for the Flare+$\alpha2.5$CNM case is greater than the value of $n_0$ as labeled in Table \ref{tab:allmodels}; in addition, the initial density for the Flare+$\alpha2.5$CNM case is also greater than the density encountered by the flare for the Flare+$\alpha0$CNM case when considering the same listed value of $n_0$ in Table \ref{tab:allmodels} (see Figures \ref{fig:M0p01_shellevol} and \ref{fig:M0p1_shellevol}). 

In general, the shock in our models starts to decelerate significantly when it reaches the radius $R_{\rm dec}$, where the flare sweeps up a mass comparable to its own mass. For the constant density Flare+$\alpha0$CNM models, this occurs at approximately
\begin{equation}
\label{eq:rdec}
    R_{\rm dec}=\left(\frac{3E}{4\pi n_0 m_p v^2}\right)^{1/3}.
\end{equation}
The value of $R_{\rm dec}$ depends on the flare energy $E$, the velocity of the shock, $v$, and the constant density of the CNM, $n_0$. For larger values of $n_0$, the shock therefore decelerates at smaller radii as the shock sweeps up the denser surrounding gas more quickly. This Sedov-Taylor solution leads to the following scaling of shock velocity with radius $R_{\rm sh}> R_{\rm dec}$: 
\begin{equation}
    v_{\rm sh} \propto R_{\rm sh}^{(-3+\alpha)/2}.
\end{equation}
Thus, the shock velocity decreases much less dramatically for the Flare+$\alpha2.5$CNM case, as seen in Figures \ref{fig:M0p01_shellevol} and \ref{fig:M0p1_shellevol}. 

\begin{figure*}
    \centering
    \includegraphics[width=\textwidth]{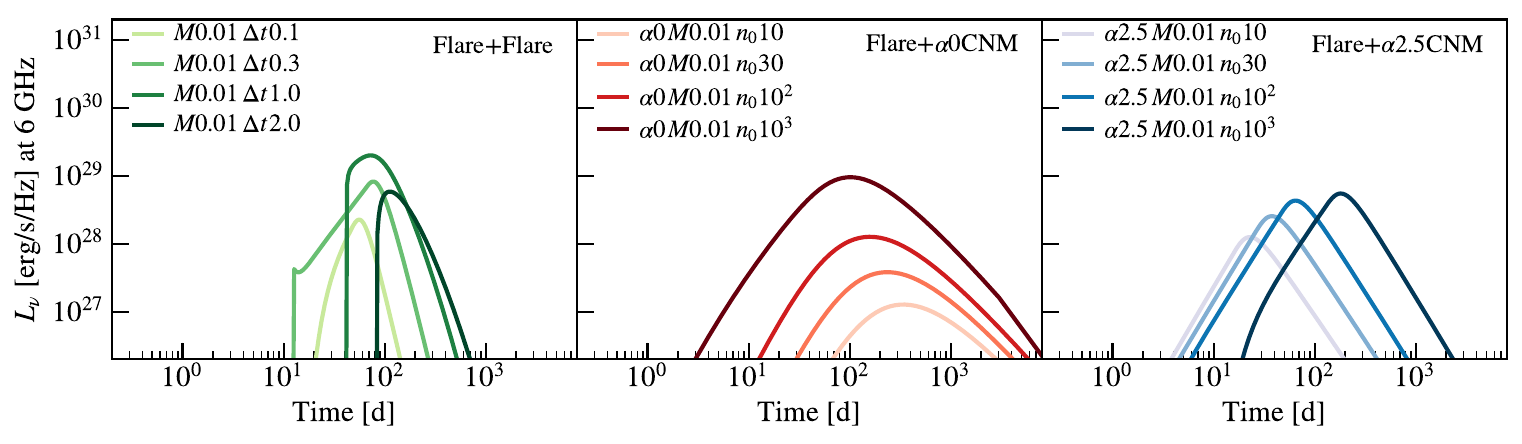}
    \includegraphics[width=\textwidth]{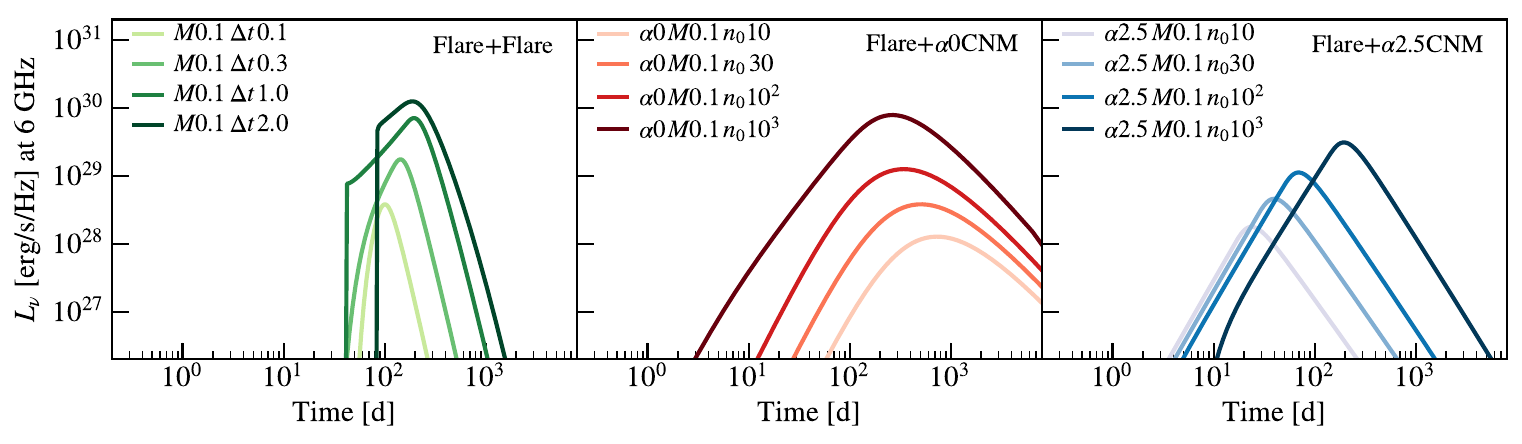}
    \caption{Light curves of radio emission at 6 GHz for the models shown in  Figure \ref{fig:M0p01_shellevol} (top row) and Figure \ref{fig:M0p1_shellevol} (bottom row).}
    \label{fig:allmodelLCs}
\end{figure*}

Interaction with the constant density CNM can slow the shock to $\lesssim 0.1c$ at radii of $\sim 10^{17}$--$10^{18}$ cm. For the Flare+$\alpha2.5$CNM models, the shock velocity tends to remain larger than $\sim 0.1c$ unless the CNM density at $10^{17}$ cm is $n_0\gtrsim 10^3\, \mathrm{cm}^{-3}$. Overall, the velocities in our models are consistent with the inferred outflow properties from \cite{cendes2024} and \cite{goodwin2025arXiv}, which estimate lower limits to the outflow speeds in the range of $\approx 0.01$--$0.1c$.

\begin{figure*}
    \includegraphics[width=\textwidth]{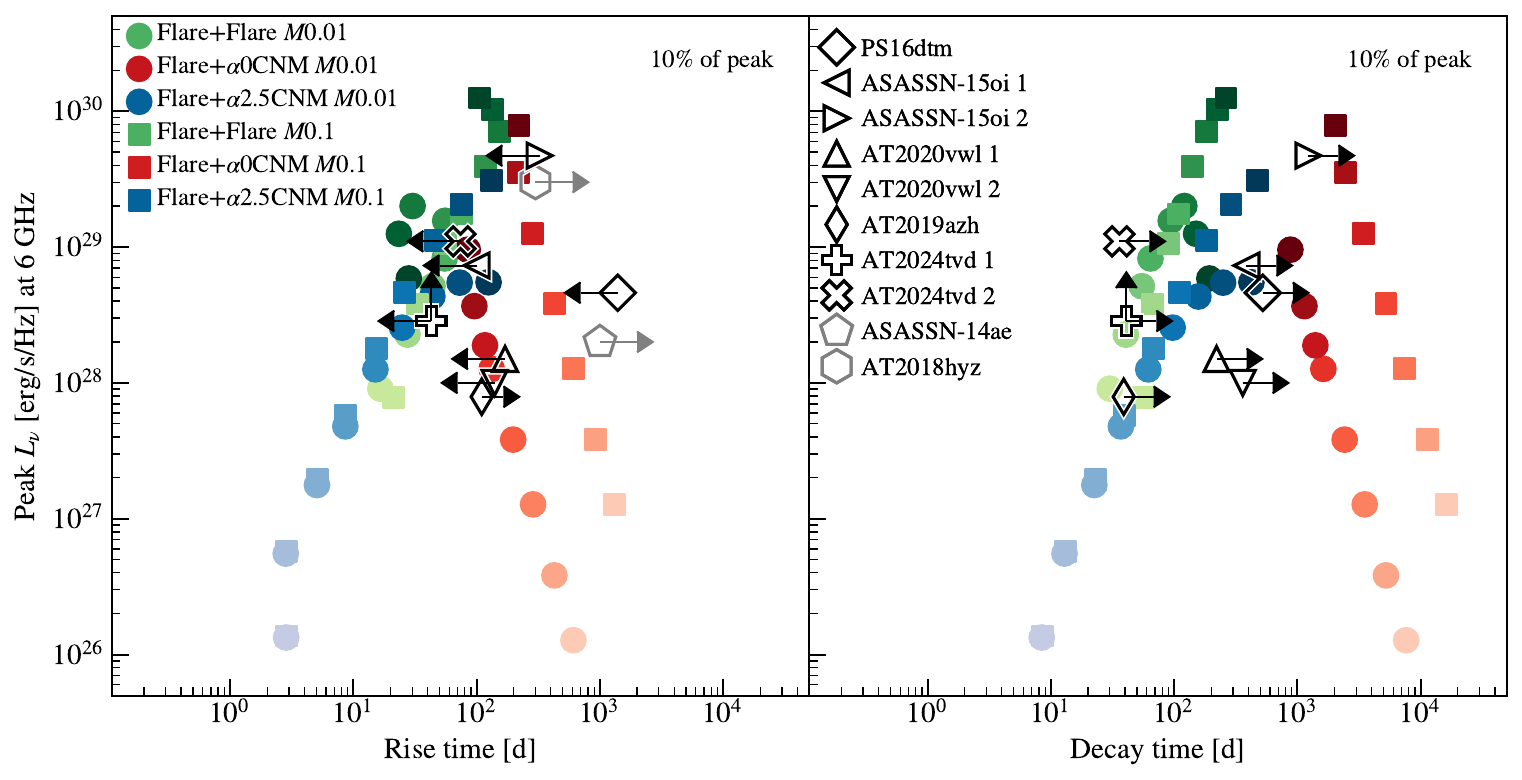}
    \caption{Peak properties of our model light curves at 6 GHz. The left panel shows the peak luminosity versus the rise time, and the right panel shows the peak luminosity versus the decay time; these are defined in Section \ref{sec:peakprops}. The scatter points correspond to Flare+Flare in green, Flare+$\alpha2.5$CNM in blue, and Flare+$\alpha0$CNM in red; circles are $M_{\rm fl}=0.01\, M_{\odot}$ models, while squares are $M_{\rm fl} =0.1\, M_{\odot}$ models. Darker shades indicate higher CNM density for red and blue points, or larger $\Delta t$ for green points. To show the extent of parameter space, we include more models than are listed in Table \ref{tab:allmodels} or shown in Figures \ref{fig:M0p01_shellevol}--\ref{fig:allmodelLCs}. 
    For the red Flare+$\alpha0$CNM models, the rise/decay times trend shorter for darker shaded points because the peak timescales are typically set by the Sedov timescale (Equation \ref{eq:sedovtime}). For the blue Flare+$\alpha2.5$CNM and green Flare+Flare models, the peak timescales are set by when $\tau_{\rm SSA}\rightarrow 1$, so darker shades trend towards longer rise/decay times.
    Estimates of the peak flux and rise or decay time are also shown for a selection of observed TDEs, listed in the legend in the right panel. Lower limits indicate events where the peak flux is not constrained, and upper or lower limits to the rise and decay times are also shown for events where the time to reach 10\% of the peak flux is uncertain; however, the length of the error bar is not representative of the actual uncertainty and is shown for illustrative purposes only. }
    \label{fig:rise_decay_times}
\end{figure*}

\subsection{Radio light curves}

In Figure \ref{fig:allmodelLCs}, we show the light curves  $L_{\nu}$ at 6 GHz as a function of time in days, for the models listed in Table \ref{tab:allmodels}. For the Flare+Flare models, time is measured from the time of the launch of the second outflow. In general, the Flare+Flare models lead to more short-lived light curves that rise and decline from peak on timescales of $\sim 10$s--$100$s of days. In contrast, the collision of a flare with a constant-density CNM produces light curves that can be sustained over $\sim 10^2$--$10^3$ d. For a power-law density CNM with $\alpha=2.5$, the light curves peak on shorter timescales and achieve lower peak fluxes for lower CNM densities, but last longer and are brighter as the normalization of the CNM density increases.  Throughout all models, more massive flares produce brighter light curves overall (second row of Figure \ref{fig:allmodelLCs}).  

\subsubsection{Collision of two flares}
The peak of each light curve for the Flare+Flare calculations is set by when the SSA optical depth decreases to unity, $\tau_{\rm SSA}\rightarrow1$. For smaller values of $\Delta t$, the flare densities are larger when the collision occurs, so the free-free optical depth can also be large. In that case, the rise to peak is affected by both FFA and SSA until $\tau_{\rm FFA}\rightarrow 1$, on timescales of $\sim 10$s of days. 

For large values of $\Delta t$, the flare densities can be low enough that the material ahead of the shock is initially optically thin to SSA. This leads to a transient feature visible in the early part of the light curve, during which the population of non-thermal electrons increases as electrons are injected into the shocked region. The unabsorbed synchrotron flux and the SSA absorption coefficient both rise during this phase, which ends once the SSA optical depth rises above unity and begins to mediate the rise of the light curve. 
The appearance of this initial sudden rise is subject to uncertainties in our assumptions for the flare density profile, which currently exhibits sharp cutoffs at $v_{\rm min}$ and $v_{\max}$. Furthermore, these light curves are calculated assuming that the flares do not interact with any material until they collide, while in reality interaction with CNM like in our other models would also produce some emission that would affect the total light curve during the sharp rise.
Regardless, these features last for only $\lesssim 2$d in our models.

\begin{figure}
    \centering
    \includegraphics[width=\columnwidth]{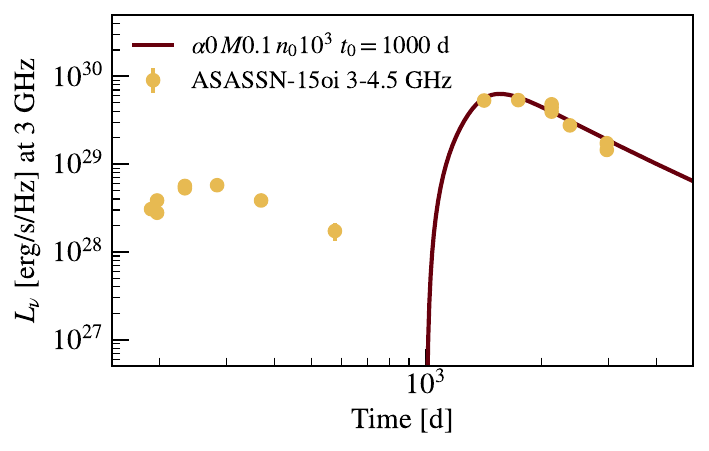}
    \includegraphics[width=\columnwidth]{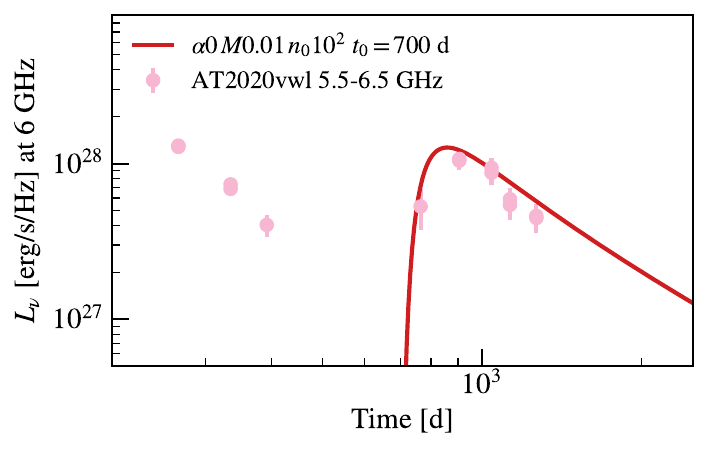}
    \caption{Top: Comparison of a selected model with radio emission from the TDE ASASSN-15oi. The model light curves shown are at 3 GHz. Bottom: Comparison of a selected model with radio emission from the TDE AT2020vwl. The model light curves shown are at 6 GHz.}
    \label{fig:ASASSN15oi_LC}
\end{figure}

\subsubsection{Collision of a flare with the CNM}
Our Flare+$\alpha0$CNM models are typically in the regime where the peak time can be approximated by the Sedov time, when the flare has swept up comparable mass to its own at a radius  $R_{\rm dec}$ \citep{Nakar11}. The corresponding time $t_{\rm dec}$ where the flare begins to decelerate is
\begin{equation}
\label{eq:sedovtime}
    t_{\rm dec} = \frac{R_{\rm dec}}{v}.
\end{equation}
For the Flare+$\alpha0$CNM models, the peak timescales of the light curves are in agreement with $t_{\rm dec}$ calculated using $v\approx v_{\rm max}$ and Equation \ref{eq:rdec}. In addition, estimates of the peak flux based on the kinetic energy of the flare and the CNM density are consistent with the peak properties of most of the Flare+$\alpha0$CNM models \citep{Nakar11}. 
The only exceptions are the two models with $n_0=10^3\, \mathrm{cm}^{-3}$, where the peak is mediated by when the SSA optical depth drops to unity, $\tau_{\rm SSA}\rightarrow 1$. 
For these models with very high density $n_0$, the flare deceleration time (Equation \ref{eq:sedovtime}) is much shorter than the peak timescale. The decrease in $\tau_{\rm SSA}$, and accordingly the peak timescale, is primarily determined by the declining B field during the deceleration of the shock velocity ($B\propto v_{i}$, Equation \ref{eq:Bfield}).
The resulting peak is slightly dimmer and delayed from the prediction based on the Sedov time (assuming $\tau_{\rm SSA}<1$), but still consistent with the expectations of \cite{Nakar11} for the case when SSA is important.

While for the Flare+$\alpha0$CNM models, the peak timescale trends inversely with the normalization of the CNM density $n_0$, the Flare+$\alpha2.5$CNM light curves instead peak later as $n_0$ increases. This effect occurs because the peak timescales of the Flare+$\alpha2.5$CNM models are set by when  $\tau_{\rm SSA}\rightarrow 1$, similar to the Flare+Flare models and the highest density Flare+$\alpha0$CNM models. 


\subsection{Peak timescale and flux}
\label{sec:peakprops}
Overall, the luminosities of our model light curves span the range of the observed delayed radio emission seen in, e.g., \cite{goodwin2025arXiv} and \cite{cendes2024}, both of which show light curves in the range of $\nu L_\nu\approx 10^{36}$--$10^{39}$ erg/s. 
Where possible, we would like to quantify whether the available observations can distinguish between differences among the three categories of models we have explored. For events where the data show a peak in the radio emission, we estimate the peak flux and timescale of the light curve's rise to peak and compare to those of our models. 

Figure \ref{fig:rise_decay_times} shows how our models lie in the parameter space of the peak luminosity $L_\nu$ at 6 GHz versus the rise time (left) and versus the decay time (right).  Here, rise time is defined as the time for the light curve to rise from 10\% of the peak luminosity to the peak of the light curve, and decay time is similarly defined as the time to decline from the peak of the light curve to 10\% of the peak.  Square points represent more massive flares with $M_{\rm fl}=0.1\, M_{\odot}$, and circles denote $M_{\rm fl}=0.01\, M_{\odot}$ models. 

More models have been added to fill in the parameter space in Figure \ref{fig:rise_decay_times} than in listed in Table \ref{tab:allmodels} and shown in Figures \ref{fig:M0p01_shellevol}--\ref{fig:allmodelLCs}, so the color shading does not correspond directly to the legends in Figures \ref{fig:M0p01_shellevol}--\ref{fig:allmodelLCs}. Nevertheless, lighter shaded points still represent less dense CNM (smaller $n_0$) or shorter $\delta t$. For the Flare+$\alpha0$CNM models in red, $n_0$ ranges from $10^0$--$10^3\, \mathrm{cm}^{-3}$; for the Flare+$\alpha2.5$CNM models in blue $n_0$ ranges from $3\times10^{-2}$--$10^3\, \mathrm{cm}^{-3}$; and for the Flare+Flare models in green, $\Delta t$ ranges from $0.05$--$2$ yr.

Figure \ref{fig:rise_decay_times} shows clear trends in the peak flux versus rise time and peak flux versus decay time for each type of collision model. 
The Flare+$\alpha0$CNM models tend towards shorter rise and decay times as the CNM density increases. The opposite trend is visible for the Flare+$\alpha2.5$CNM models, for which the peak flux, rise time, and decay time are all positively correlated with each other. The Flare+Flare models also occupy a similar region in the parameter space of peak flux versus rise time as the Flare+$\alpha2.5$CNM models, but the right panel shows that the brightest Flare+Flare models decay on shorter timescales. 

Overall, the rise and decay timescales are systematically longer for the Flare+$\alpha0$CNM models than for the other two types of collision models, with light curves that can decay over $\sim10^3$ d. The rise times are shorter than the decay times, on the order of $\sim 10^2$ d for the brightest light curves. 
The models with the shortest rise and decay times of $\lesssim 10$--$100$ d are from the Flare+$\alpha2.5$CNM models with the lowest CNM densities $n_0\lesssim 10^2 ~\mathrm{cm}^{-3}$, but these also correspond to dimmer light curves overall. 
Notably, the Flare+Flare models are able to produce light curves with bright peak flux $\gtrsim 10^{29}~\mathrm{erg}\, \mathrm{s}^{-1}\, \mathrm{Hz}^{-1}$ that evolve rapidly, on $\lesssim 100$ d timescales. At higher CNM densities, the rise times for the Flare+$\alpha2.5$CNM models are similar to the Flare+Flare models, but the decay times tend to be a factor of a few--10 times longer. 

\begin{figure}
    \centering
    \includegraphics[width=\columnwidth]{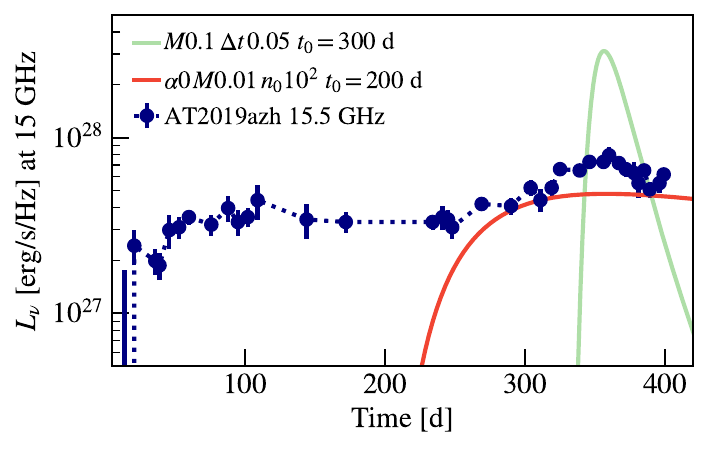}
    \caption{Comparison of two flare collision models with AT2019azh. Data is shown at 15.5 GHz, model light curves are at 15 GHz. The chosen models exhibit the most similar peak properties to AT2019azh according to Figure \ref{fig:rise_decay_times}. }
    \label{fig:AT2019azh}
\end{figure}

\begin{figure}
    \centering
    \includegraphics[width=\columnwidth]{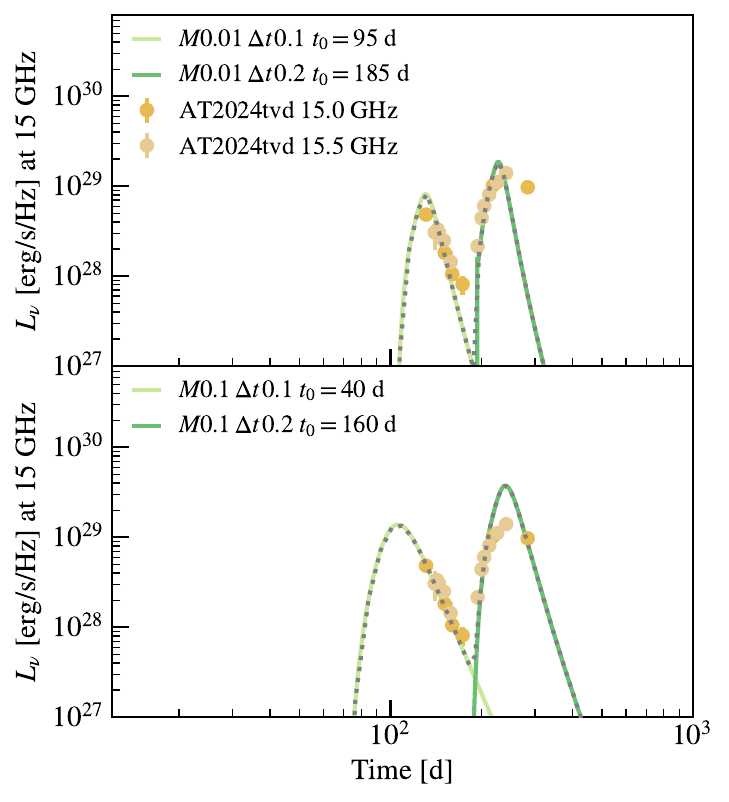}
    \caption{Comparison of a selection of models with  AT2024tvd. The model light curves shown and listed in the legend are at 15 GHz. The gray dotted line in each panel shows the sum of both model light curves.}
    \label{fig:AT2024tvd_LC}
\end{figure}

Figure \ref{fig:rise_decay_times} also includes estimates of the rise time, decay time, and peak flux for several events that display well-constrained peaks, shown as black outlined shapes. 
These values are estimated from the available epochs in the data, so the location of the scatter points for real TDEs in Figure \ref{fig:rise_decay_times} should be regarded as highly approximate, at within a factor of a few.
We discuss the comparison between these events and our models below.  To align our model light curves with the delayed radio emission observed in these events, we can shift the initial time $t_0$ of the model light curve to a nonzero value; the magnitude of the shift is indicated by $t_0$ at the top of each panel in Figures \ref{fig:ASASSN15oi_LC}--\ref{fig:AT2024tvd_LC}.

\textit{\textbf{PS16dtm}}: The event PS16dtm (square diamond in Figure \ref{fig:rise_decay_times}) peaks at $\approx 4\times 10^{28}\, \mathrm{erg}\, \mathrm{s}^{-1}\, \mathrm{Hz}^{-1}$ with a rise to peak of $\lesssim 10^3$ d and a decay from peak over $\gtrsim 500$ d, beyond which the decline is not fully constrained \citep{cendes2024}. Compared to the peak properties of our models, the peak flux and timescales are consistent with the collision of a $M_{\rm fl}=0.1\, M_{\odot}$ flare with constant density CNM with $n_0=30\, \mathrm{cm}^{-3}$. The top panel of Figure \ref{fig:PS16dtm_LCSED} shows the agreement between our model light curve and the peak seen in the data.

\textit{\textbf{ASASSN-15oi}}: We also compare our models to ASASSN-15oi \citep{horesh2021,hajela2025,alexander25}, which shows evidence for two delayed peaks in the radio emission. As Figure \ref{fig:rise_decay_times} indicates, the second peak (right-facing triangle in Figure \ref{fig:rise_decay_times}) appears to show similar timescales and peak flux to the collision of a flare with dense CNM. The top panel of Figure \ref{fig:ASASSN15oi_LC} shows how our model light curve at 3 GHz for an $M_{\rm fl}=0.1\, M_{\odot}$ flare colliding with $n_0=10^3\, \rm{cm}^{-3}$, constant density CNM matches the second peak well. 
We note that the accelerated electron power-law index $p$ inferred from the SED varies significantly between each epoch \citep[see e.g.][Table 2]{hajela2025}. At the second peak, the observed SEDs favor smaller values of $p <3$, whereas the models shown assume a constant $p=3$; for lower values of $p$, we would expect flatter model SEDs that are more consistent with the observations, as well as a shallower light curve decay.

However, the first radio peak (left-facing triangle in Figure \ref{fig:rise_decay_times}) is narrower than the nearest model in Figure \ref{fig:rise_decay_times} can reproduce. This earlier radio emission may originate from a mechanism other than the flare collisions we model here. For instance, \cite{hajela2025} attribute the first radio flare to collision-induced outflows from stream-stream interaction.

\textit{\textbf{AT2020vwl}}: Another event of interest is AT2020vwl \citep{goodwin2023,goodwin2025}, which also exhibits two peaks in the radio emission after the optical light from the TDE. Figure \ref{fig:rise_decay_times} shows that the peaks (upward and downward triangles) have similar peak flux to a $M_{\rm fl}=0.01\, M_{\odot}$ flare colliding with a $n_0=10^2\, \mathrm{cm}^{-3}$, constant density CNM. The bottom panel of Figure \ref{fig:ASASSN15oi_LC} shows that the model can explain the observed light curve at $\approx 6$ GHz fairly well, though the data decline a little faster than the model light curve. 
The observed SEDs are consistent with $p=3$, as assumed in our models, at all epochs.
The initial radio peak declines more steeply than the models with similar peak flux in Figure \ref{fig:rise_decay_times} predict; it may be powered by a different mechanism, such as an earlier outflow from stream-stream collisions \citep{goodwin2023}.

\textit{\textbf{AT2019azh}}: \cite{sfaradi2022} suggest the presence of a late peak in the radio emission at 15.5 GHz for the TDE AT2019azh. When we estimate the properties of this peak (thin diamond in Figure \ref{fig:rise_decay_times}), they fall in a regime of parameter space that is similar to constant density Flare+$\alpha0$CNM models for the rise time, but more similar to the Flare+$\alpha2.5$CNM models or Flare+Flare models for the decay time. A caveat is that the model properties plotted in Figure \ref{fig:rise_decay_times} assume a light curve at 6 GHz, whereas the peak observed for AT2019azh is at 15.5 GHz. 

Figure \ref{fig:AT2019azh} demonstrates that no single model light curve (at 15 GHz) can simultaneously fit both the rise and decay of the light curve bump seen in AT2019azh at around $\sim 300$--$400$ d. On the one hand, the red Flare+$\alpha0$CNM model rises on the same timescale, but it is too flat compared to the variability seen at $\sim300$--400 d. On the other hand, the green Flare+Flare model declines more rapidly, but also rises too sharply compared to the data. The 15 GHz light curve is also brighter than what is seen in AT2019azh; based on the trend in Figure \ref{fig:rise_decay_times}, a dimmer Flare+Flare model could match the brightness of late peak for AT2019azh, but would likely decay too quickly. A steeper flare density profile with $s > 0.5$ might be relevant to this event, as this assumption would lead to slower shock velocities and therefore more slowly-evolving, fainter light curves.

\textit{\textbf{AT2024tvd}}: Finally, we compare our models to AT2024tvd \citep{yao2025,sfaradi2025}, which displays two radio peaks that rise and decline very steeply with time. The Flare+Flare models are good candidates to explain such sharp peaks that are also relatively bright (plus sign and cross in Figure \ref{fig:rise_decay_times}). Figure \ref{fig:AT2024tvd_LC} shows some examples of how our Flare+Flare models compare to the peaks seen in AT2024tvd at 15 GHz. While we have not performed detailed modeling of the scenario that would produce each peak, we see from Figure \ref{fig:AT2024tvd_LC} that the morphologies of the narrow peaks seen in each model light curve strongly resemble the steep rise and fall of the observed two peaks. 
A scenario to explain the two peaks with these models would involve two episodes of Flare+Flare interaction occurring $\sim 100$--$200$ d apart, due to multiple flares ejected from the disk that result in multiple flare collisions.
One caveat is that the inferred electron spectrum power-law index from the observed SEDs ($p \sim 2.1$--$2.2$) is lower than the models shown in this work, which assume $p=3$. The models therefore predict steeper SEDs than are observed at high frequencies, though this discrepancy could be alleviated with $p<3$.  In addition, the model light curve with $p<3$ would decline less rapidly, which could improve the comparison between the $M_{\rm fl}=0.01\, M_{\odot}$ models and the observations.


Notably, the delay time we assume for the second flare $M=0.01\, M_{\odot}$ case is consistent with the outflow launch time inferred by \cite{sfaradi2025}.
An interpretation of the two peaks as arising from flare collisions may also be supported by the peculiar off-nuclear nature of this TDE, which may indicate a lower ambient density. This would imply relatively stronger contribution from emission due to our Flare+Flare models compared to emission from the collision of a flare with the CNM.
We do caution that at these early times of $\sim 100$ d, inverse Compton cooling is important, which we have not accounted for in our models, and this may lessen the radio emission; other assumptions for the shock microphysical parameters $\epsilon_B$ or $\epsilon_e$ would also affect the model light curves. Such an effect may be particularly relevant for the second peak, which appears dimmer than our  $M_{\rm fl}=0.1\, M_{\odot}$ model.

\begin{figure*}
    \centering
    \includegraphics[width=\textwidth]{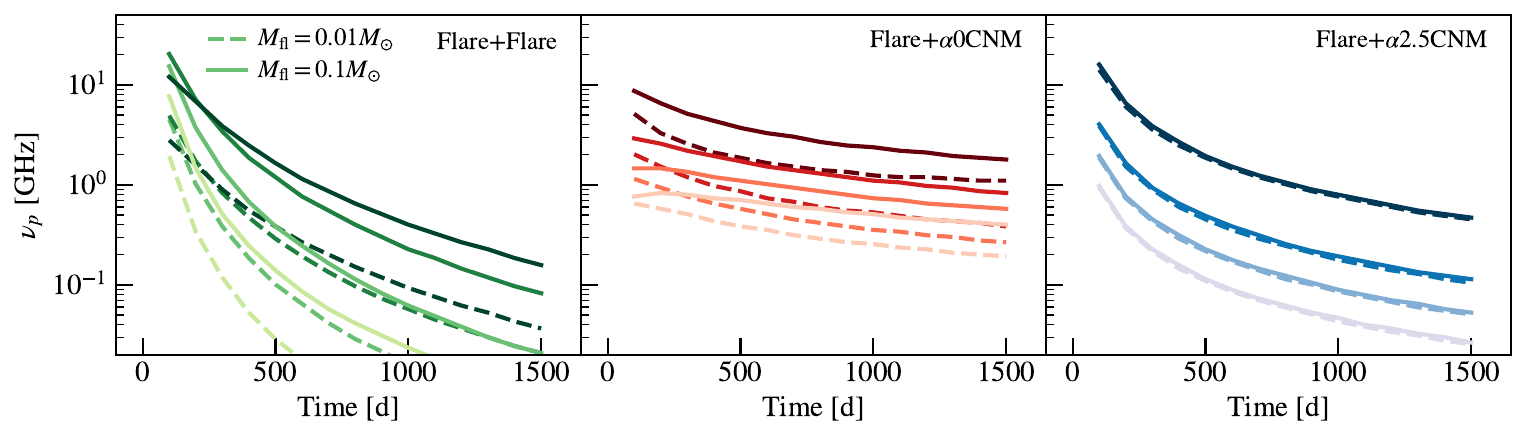}]
    \caption{
    Evolution of the peak frequency of the SED for the models shown in Figure \ref{fig:M0p01_shellevol} (dashed) and Figure \ref{fig:M0p1_shellevol} (solid). Colors correspond to the same models as in the legends of Figures \ref{fig:M0p01_shellevol} and \ref{fig:M0p1_shellevol}.
    }
    \label{fig:allmodelnupeaks}
\end{figure*}

\begin{figure}
    \centering
    \includegraphics[width=\columnwidth]{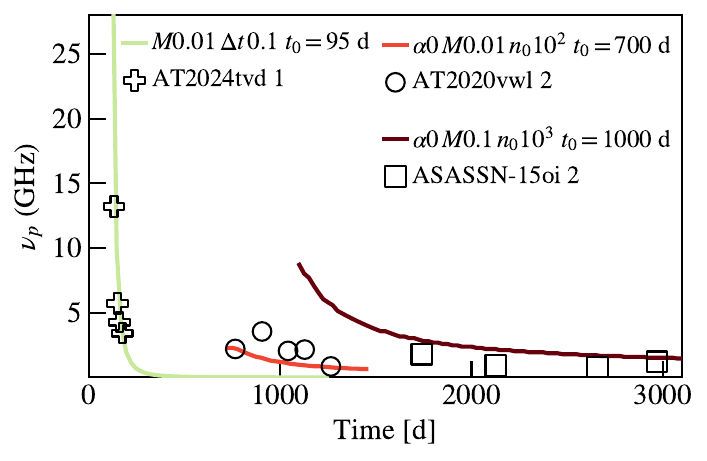}
    \caption{
    Comparison of the peak frequency evolution of our model SEDs with that of selected TDEs. The pairs of models and events are taken from Figures \ref{fig:ASASSN15oi_LC} and \ref{fig:AT2024tvd_LC}.
    }
    \label{fig:nupeak_compare}
\end{figure}

\begin{figure*}
    \centering
    \includegraphics[width=\columnwidth]{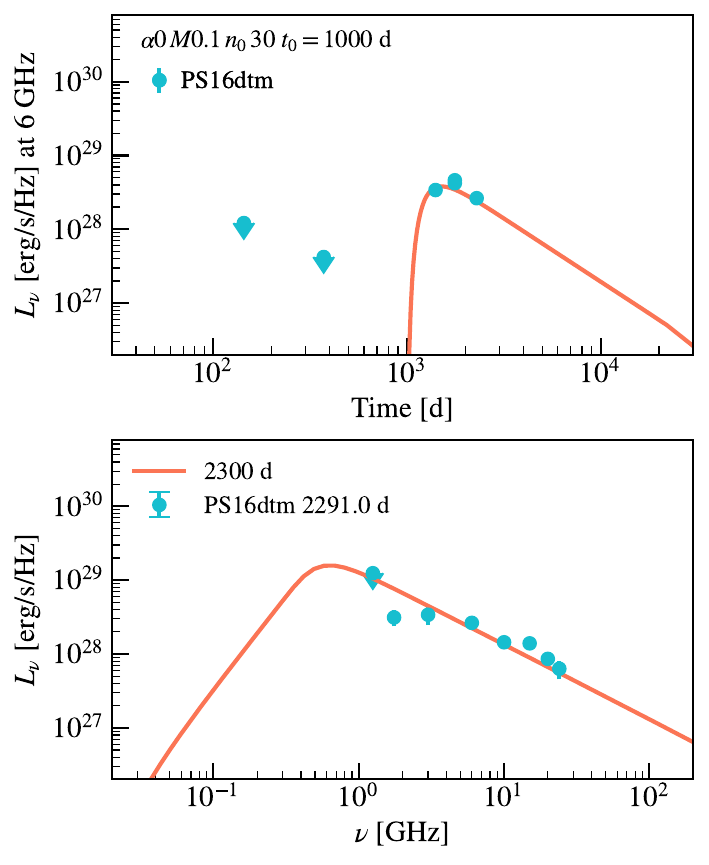}
    \includegraphics[width=\columnwidth]{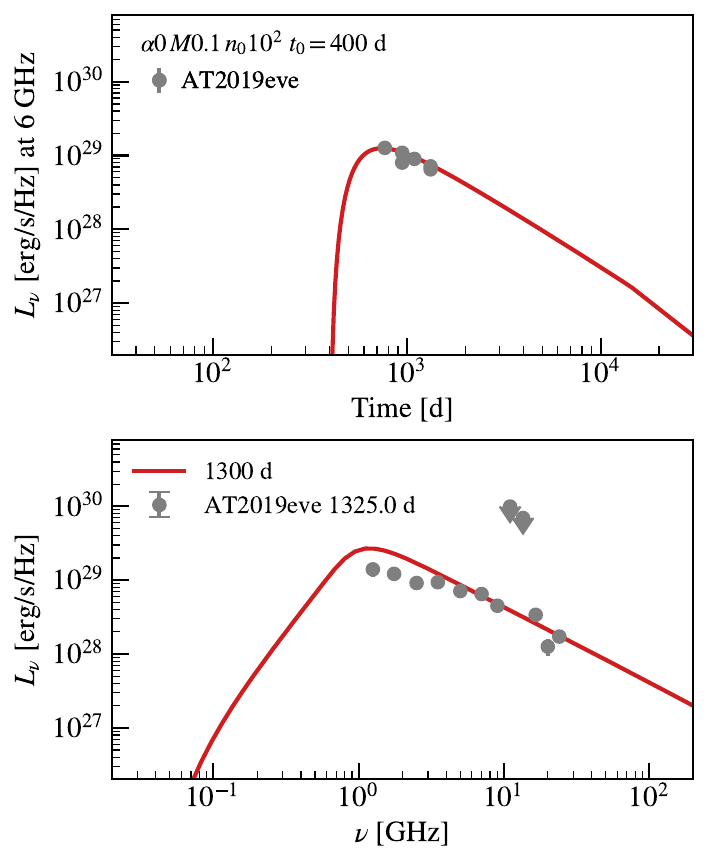}
    \caption{Comparison of Flare+$\alpha0$CNM models with radio emission from PS16dtm (left) and AT2019eve (right). For each event, we show the light curve at 6 GHz and the SED at the epoch listed in the legend.}
    \label{fig:PS16dtm_LCSED}
\end{figure*}

In general, Figure \ref{fig:rise_decay_times} shows that our brightest models tend to have rise times on the order of $\sim 100$ d. As a result, our models struggle to explain events that exhibit a rise over $\sim 1000$s of days, but are also relatively bright; we show two such events as gray outlined shapes in Figure \ref{fig:rise_decay_times}. For instance, the event ASASSN-14ae (gray pentagon) is noted to have quite a steep increase in its flux, but it is still rising on timescales of $\gtrsim 1000$ d to a peak flux $\gtrsim 10^{28} \mathrm{erg}\, \mathrm{s}^{-1}\, \mathrm{Hz}^{-1} $ that has not yet been constrained \citep{cendes2024}. While Flare+$\alpha0$CNM models with lower density CNM can rise on timescales of up to $\sim 1000$ d, their peak $L_{\nu}$ values are far too low to explain such an event. A similar discrepancy is evident for AT2018hyz (gray hexagon), whose radio emission is detected while rising rapidly on timescales of $\gtrsim 300$ d to a bright peak luminosity of at least $\gtrsim 2\times 10^{39} \mathrm{erg}\, \mathrm{s}^{-1}\, \mathrm{Hz}^{-1}$ \citep{cendes2022}. Though this peak flux can be achieved by our brightest models, the rise of AT2018hyz is actually shallower than the light curves of those models.

In summary, we find that the Flare+Flare models are well suited to explain events showing bright, delayed radio emission that evolves quickly on $\sim 100$ d timescales. Other events that evolve on longer timescales of $\sim 10^3$ d may be more indicative of a flare interacting with constant density CNM. Our Flare+$\alpha2.5$CNM models are likely to explain events that rise on short timescales, but decay on timescales that are longer than the Flare+Flare models by a factor of a few. While none of the comparisons with events exhibiting well-constrained peaks included the Flare+$\alpha2.5$CNM models, we show some examples of events that may fit well with Flare+$\alpha2.5$CNM models in Section \ref{sec:sedcomparison}. 

\subsection{Peak frequency of the SED}

Another commonly observed property of the radio emission in TDEs is the peak frequency of the SED, $\nu_p$, and how $\nu_p$ changes across different epochs. We show the evolution of $\nu_p$ as a function of time in Figure \ref{fig:allmodelnupeaks}. 
For the simplest case of constant density material and a peak set by synchrotron self-absorption, the peak frequency would be expected to decline approximately as $\nu_p \propto t^{-1}$, once the flare has swept up a mass of material comparable to its own \citep{Nakar11}.
The Flare+$\alpha0$CNM models in the middle column follow a similar trend, with expected deviations since our velocity evolution is more sophisticated than that assumed in \cite{Nakar11}. We see that the density profile of the exterior material influences the evolution of the peak frequency, as the steeper density profiles of the Flare+Flare and Flare+$\alpha2.5$CNM models exhibit more rapid decreases in $\nu_p$ with time.

In Figure \ref{fig:nupeak_compare}, we compare the evolution of $\nu_p$ in our models to the inferred SED peak frequency for a few events. The observed evolution of $\nu_p$ for AT2024tvd from $\approx13$ GHz to $\approx 3$ GHz in $\approx 40$ d is well-reproduced by the rapid decline in $\nu_p$ exhibited by our Flare+Flare model. This is true of the $M_{\rm fl}=0.1\, M_{\odot}$ model as well. For AT2020vwl and ASASSN-15oi, the observed values of the peak frequency are typically $\lesssim 5$ GHz, which is generally consistent with the lower values of $\nu_p$ that are typical of the Flare+$\alpha$0 CNM models. We note that the observed SEDs for ASASSN-15oi in particular required a multi-component power-law fit to reproduce the multiple breaks \citep{hajela2025}, so our model SEDs may not fully capture the physics influencing the SEDs of certain TDEs.

\subsection{Light curve and SED comparisons}
\label{sec:sedcomparison}

\begin{figure*}
    \centering
    \includegraphics[width=\columnwidth]{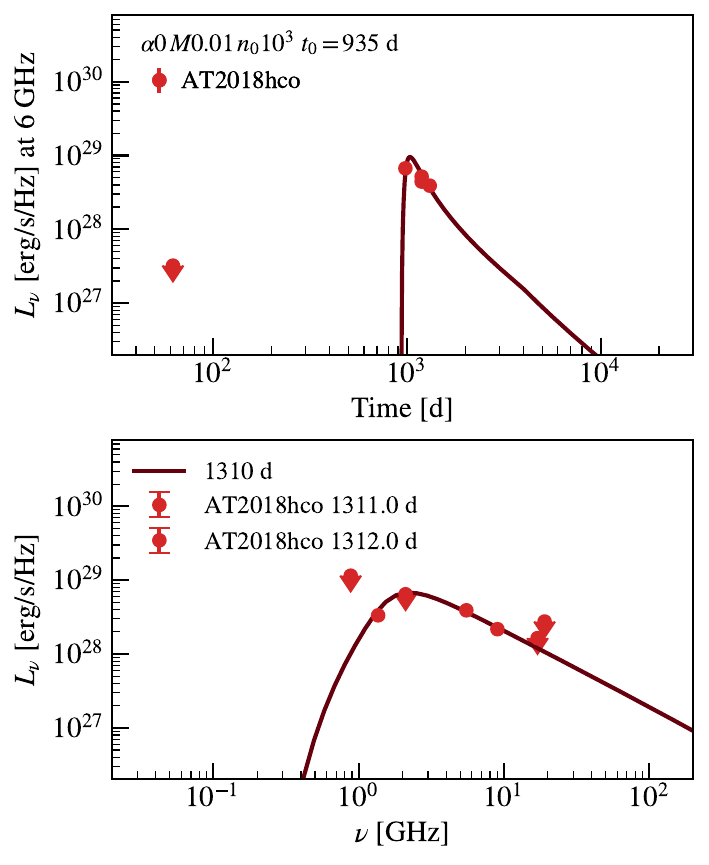}
    \includegraphics[width=\columnwidth]{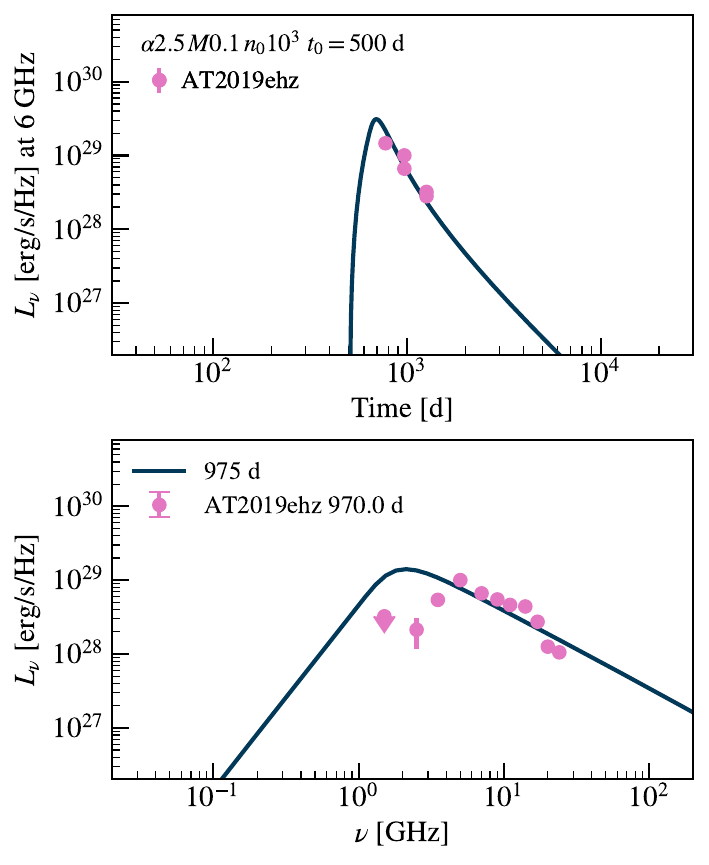}
    \caption{Comparison of a Flare+$\alpha0$CNM model with AT2018hco (left) and a Flare+$\alpha2.5$CNM model with radio emission from AT2019ehz (right). For each event, we show the light curve at 6 GHz and the SED at the epoch listed in the legend.}
    \label{fig:otherexamples}
\end{figure*}

\begin{figure*}
    \centering
    \includegraphics[width=\textwidth]{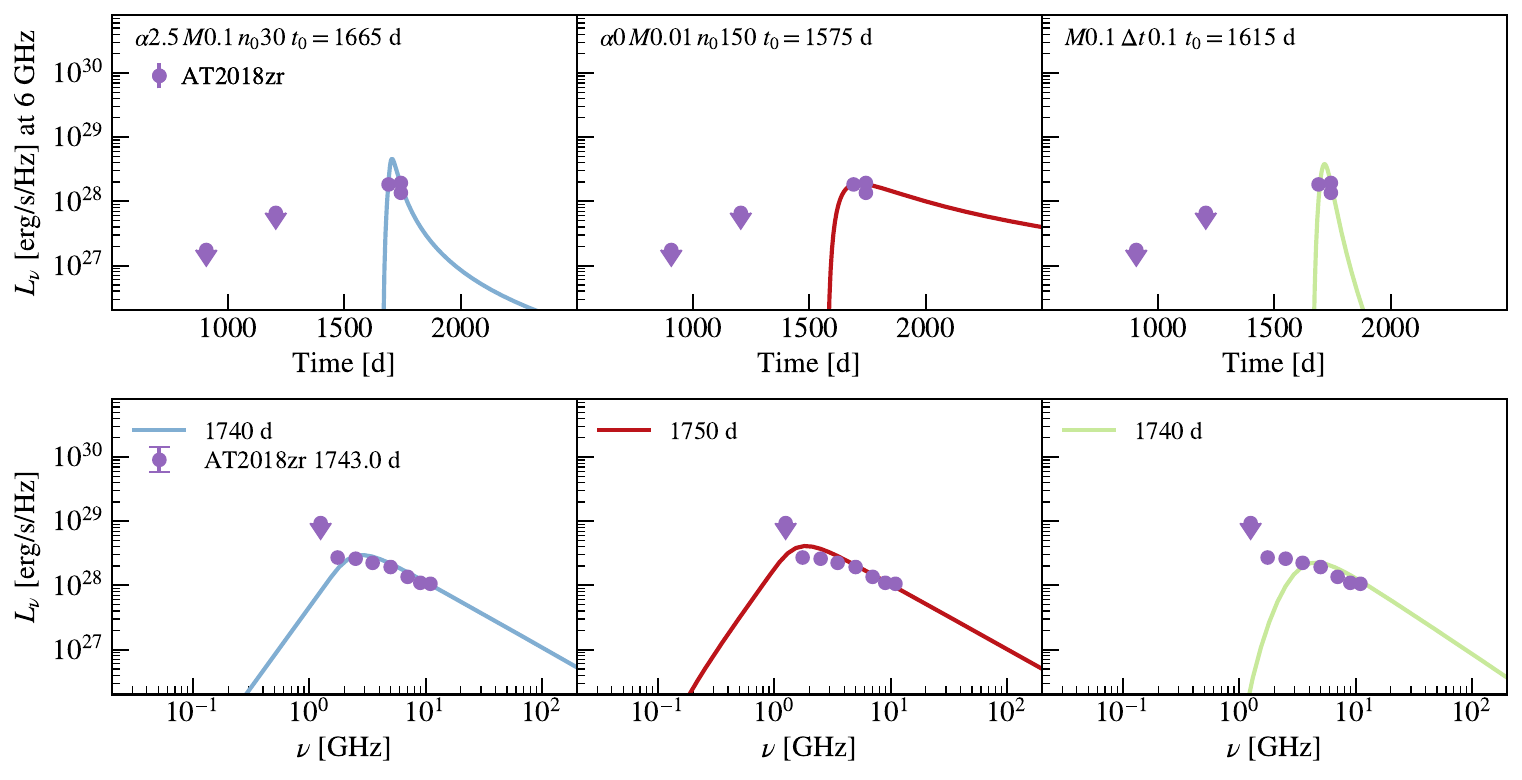}
    \caption{Comparison of model radio emission from a selection of models with radio emission from the TDE AT2018zr. Top are light curves at 6 GHz, and bottom are SEDs at the epoch listed in the legend.}
    \label{fig:AT2018zr_LCSED}
\end{figure*}

\begin{figure*}
    \centering
    \includegraphics[width=\textwidth]{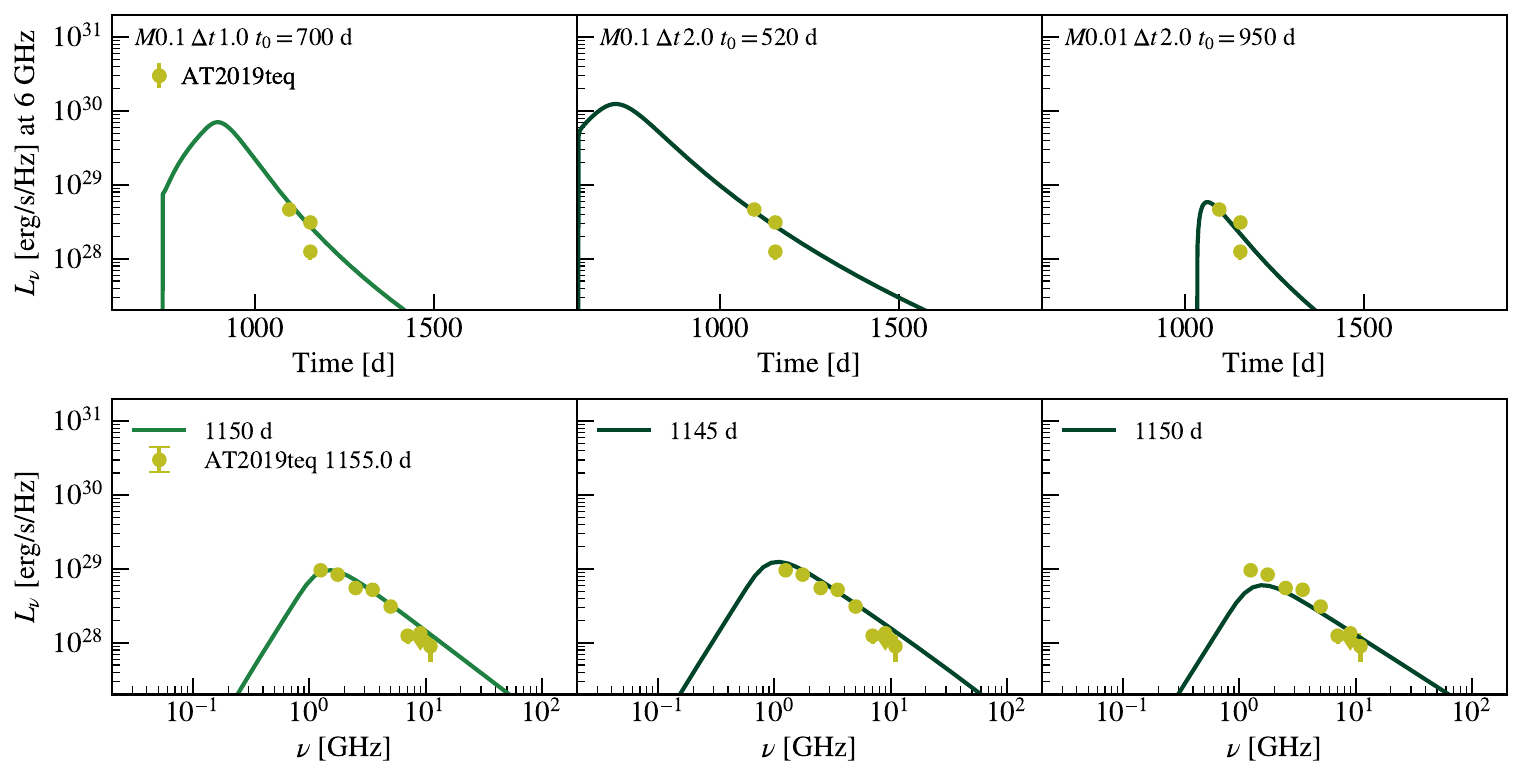}
    \caption{Comparison of model radio emission from a selection of the Flare+Flare models with radio emission from the TDE AT2019teq. Top are light curves at 6 GHz, and bottom are SEDs at the epoch listed in the legend.}
    \label{fig:AT2019teq_LCSED}
\end{figure*}

Figures \ref{fig:PS16dtm_LCSED}--\ref{fig:AT2019teq_LCSED} 
show examples of comparisons between the light curves and SEDs of our models and a subset of TDEs (each from \citealt{cendes2024}). The name of the model shown in each column is listed in each panel in the top row (see Table \ref{tab:allmodels}). As in Section \ref{sec:peakprops}, the time shift $t_0$ of each model light curve is indicated at the top of each panel as well. The luminosity of events from \cite{cendes2024} is shown at 6 GHz where available, or from 5--7 GHz otherwise.

\textit{\textbf{PS16dtm}}:
For PS16dtm, the apparent peak is similar to that of the light curve produced by one of our Flare+$\alpha0$CNM models, with $M_{\rm fl}=0.1\, M_{\odot}$, $n_0 = 30\, \mathrm{cm}^{-3}$ and a delay time of $10^3$ d (Figure \ref{fig:PS16dtm_LCSED}, left). This is consistent with our expectation from the peak properties as seen in Figure \ref{fig:rise_decay_times}. Though the peak of the observed SED is not constrained, the model SED at the time of that observed epoch appears to follow a similar decline, which also resembles the fits from modeling of PS16dtm by \cite{cendes2024}. 

\textit{\textbf{AT2019eve and AT2018hco}}:
Not all events with delayed radio emission resolve the peak of the light curve. For instance, our Flare+$\alpha0$CNM models fit well with the available light curve and SED data for the events AT2019eve (Figure \ref{fig:PS16dtm_LCSED}, right) and AT2018hco (Figure \ref{fig:otherexamples}, left). The available light curve of AT2019eve appears to fit well with the decline of a Flare+$\alpha0$CNM model with a larger flare mass. For AT2018hco, the model shown is for a lower flare mass, and it fits the data if the peak of the light curve occurs in a gap between consecutive epochs. Yet more data that captures the rise to peak could refine, or even potentially negate, these comparisons. 

\textit{\textbf{AT2018zr}}:
For some events, the scarcity of data at different epochs leaves the interpretation open to many different models. 
For example, AT2018zr bears resemblance to the light curves of three different models, each requiring a delay time of $\sim 1600$ d (Figure \ref{fig:AT2018zr_LCSED}). 
With very few epochs in the light curve, the available data of AT2018zr is plausibly reproduced by model light curves with quite disparate morphologies, from the very narrow peaks of the Flare+$\alpha2.5$CNM and Flare+Flare models, to a Flare+$\alpha0$CNM model that exhibits a long decay from peak. Future observations of AT2018zr could provide necessary constraints on which, if any, of the flare collision scenarios is best suited to explain the delayed radio emission.

\textit{\textbf{AT2019ehz}}: The available SEDs for the events from \cite{cendes2024} also contribute valuable constraints. For AT2019ehz (Figure \ref{fig:otherexamples}, right), we see an example where the light curve of a Flare+$\alpha2.5$CNM model appears to resemble the observed decline of the light curve. However, the SED of the model peaks at lower frequency than the apparent peak in the observed SED. A similar discrepancy arises in our comparison of a Flare+Flare model with AT2018zr (Figure \ref{fig:AT2018zr_LCSED}, right), where the model SED appears to peak at higher frequencies compared to the observed SED. The SED therefore provides important evidence that disfavors certain models. 

\textit{\textbf{AT2019teq}}:
Even with only a few data points for the light curve, some events show evident rapid evolution in the observed flux. The late time emission detected for AT2019teq constitutes two epochs that decline rapidly in flux over $60$ d, which our models of a flare colliding with either $\alpha=0$ or $\alpha=2.5$ CNM struggled to reproduce. However, many different parameter choices for the Flare+Flare models decay steeply through those epochs, resembling both the light curve and the 1155 d SED (Figure \ref{fig:AT2019teq_LCSED}).

The majority of the events with enough epochs to resolve the decline of the light curve demonstrate a longer decay time than the Flare+Flare models can explain on their own. As summarized in Figure \ref{fig:rise_decay_times}, a slow decline from the light curve peak instead favors the Flare+$\alpha0$CNM models, as well as some of the Flare+$\alpha2.5$CNM models with the densest CNM. Thus, many of the events we compare to in this work appear to fit well with the models of a flare colliding with CNM. However, the observed data is often sampled relatively sparsely in time and may not resolve the rapid rise and decline of the Flare+Flare models. Thus, we note that we cannot confidently rule out the underlying existence of radio emission as predicted from the Flare+Flare models. The collision of multiple flares may manifest as spikes in flux on $\lesssim 100$ d timescales, which could produce variability on top of the more slowly evolving light curves from the collision of a flare with the CNM.

Table \ref{tab:comparisonsummary} summarizes the potential agreement between our models and the selection of TDEs with delayed radio emission shown in this work, based on our analysis of the radio light curves and SEDs. 
Events whose light curves can be reproduced by our models, but whose SEDs are discrepant, are marked with an asterisk. We note that in some cases, the inconsistency of the SED can be traced to a different value for the electron spectrum power-law index, which can be addressed in future modeling. 

\section{Discussion and Conclusions}
\subsection{Comparisons between different scenarios for delayed radio emission}
Other scenarios have been invoked to explain delayed radio flares, such as collisionally induced outflows (CIO) from stream-stream collisions \citep{lu2020} and outflow-cloud interaction for an inhomogeneous CNM \citep{yang2025,zhuang2025}. Stream-stream CIO is expected to produce unbound material moving at $\sim 0.01c$--$0.1c$, also consistent with the inferred outflow properties for many events, and may also produce 5 GHz light curves in the range of $L_{\rm nu} \sim 10^{26}$--$10^{30}\, \mathrm{erg}\, \mathrm{s}^{-1}\, \mathrm{Hz}^{-1}$ \citep{lu2020}. In the case of an inhomogeneous CNM, the interaction of an outflow launched during the fallback or accretion process of the TDE may produce a delayed peak in the radio emission when it encounters a dense cloud of material located at $\sim 0.1$--$1$ pc \citep{yang2025,zhuang2025}. 
Both of the above scenarios have been invoked for the first radio flare in, e.g., ASASSN-15oi, which \cite{hajela2025} also suggest has a different physical origin from the second radio flare. This picture is consistent with our work, which finds that while the first radio peak is not well-matched by our flare collision models and likely requires a different explanation, the second radio peak is consistent with a model of a flare colliding with the CNM.

Another popular hypothesis for the delayed radio emission in TDEs is the presence of an off-axis relativistic jet, as proposed for, e.g., AT2018hyz \citep{cendes2022,matsumoto2023,sfaradi2024}. AT2018hyz was found to have sharply rising, bright radio emission delayed to $\approx3$ yr after optical discovery, which our flare collision models struggle to reproduce. Similarly, ASASSN-14ae exhibits a bright, ongoing steep rise, but delayed even later to $\gtrsim 2000$ d, which is inconsistent with our models. However, \cite{cendes2024} note that the long delay implies a low $\beta$ value for the outflow, which may preclude the explanation of a relativistic jet for ASASSN-14ae.

In some cases, delayed radio emission from TDEs has been associated with an outflow launched around the time of the maximum accretion rate of the TDE disk, which is connected to an observed peak in the X-ray luminosity \citep{goodwin2025,hajela2025,goodwin2025arXiv}. 
Similarly, the scenario in \cite{piro2025} that motivates our flare collision models would predict that the outflows are launched when the disk is in a high accretion rate state due to thermal instability. 
The possibility that outflows may relate to changes in the accretion disk is also discussed for AT2024tvd, as \cite{sfaradi2025} observe a spectral change in the X-ray emission coincident with the inferred launch time for the outflow that powers the first radio flare. Nevertheless, \cite{goodwin2025arXiv} find across a range of TDEs that outflows may be launched during sub-Eddington accretion states, super-Eddington accretion states, or may not appear coincident with either high accretion states or disk transitions at all.  

The onset of the thermal instability in the disk instability scenario is deferred until the disk transitions from being advective and geometrically thick to becoming geometrically thin \citep{piro2025}. This delayed onset of the unstable state, and therefore any instability-driven outflows, may relate to the observed delays in radio emission after the TDE. However, gaps in the temporal coverage of TDE radio observations and possible delays in the disk formation with respect to optical peak should also be considered as potential contributors to the delay times in some TDEs.

Based on our analysis in this work, the disk instability scenario may be particularly favorable to invoke for events with multiple rapidly evolving peaks in the radio. 
Disk instabilities can eject a few--tens of outflows within years of the TDE \citep{piro2025}, potentially producing multiple bright and rapidly evolving flares in the radio as these outflows collide. The resulting light curve will appear quite unique, with both steeply rising and decaying emission \citep[e.g.,][]{sfaradi2025}. A potential signature of flare interaction could be rapid light curve evolution on timescales of $\sim$tens of days due to the flare-flare collisions, which manifests in tandem with emission that decays on $\sim$100 day timescales from the collision of a flare with the dense CNM.

Observations of the TDE at UV wavelengths can also help constrain whether disk instabilities underlie the outflows producing delayed radio emission. \cite{piro2025} demonstrate that outflows may be linked to disk activity as the disk cycles through low and high states of accretion, with the relatively steady low state particularly bright in the UV (up to $\sim 10^{42}$ erg s$^{-1}$). The luminosity of the disk is expected to rise further to $\gtrsim 10^{42}$ erg s$^{-1}$ in the UV during the high state, when the outflow is launched. While the high state lasts only a couple days, such bright, brief UV emission would be a unique signature of disk instability, potentially detectable with high-cadence UV surveys (e.g. ULTRASAT, \citealt{shvartzvald2024}). 
Given the inferred launch times for delayed outflows powering late-time radio emission, if monitoring of the disk were to reveal UV variability directly preceding the outflow launch time, this could indicate a potential disk instability origin for those delayed outflows.

\begin{table}
\hspace{-2em}\begin{tabular}{c|c|c}
    \hline
    Flare+Flare   & Flare+$\alpha2.5$CNM & Flare+$\alpha0$CNM \\
    \hline
    AT2024tvd*, & AT2018zr, & PS16dtm, \\
    AT2019teq, & AT2019ehz* & AT2019eve, \\
    AT2018zr* & & ASASSN-15oi*, \\
    & & AT2018hco, \\
    & & AT2018zr, \\
    & & AT2020vwl \\
\end{tabular}
   \caption{Summary of which of the TDEs shown in this work are favored by each type of flare collision model. Asterisks (*) are shown next to events with light curves that are consistent with our models, but SEDs that are inconsistent under the assumptions of this work (e.g., a constant electron spectral index of $p=3$). }
\label{tab:comparisonsummary} 
\end{table}

\subsection{Model Uncertainties}
\cite{piro2025} predict flare masses $M_{\rm fl} \sim 10^{-3}$--$10^{-1}\, M_{\odot}$, with disks around lower black hole masses experiencing less time between mass ejections (e.g. smaller $\Delta t$  between flares), and larger black hole masses able to eject more massive flares. The black hole masses for the events we compare to in this work range from $M_{\rm BH} \sim 10^6$--$10^7\, M_{\odot}$. According to the results from \cite{piro2025}, this range of black hole masses may eject less mass from thermal instabilities than the largest values of $M_{\rm fl}$ that we use in our comparisons with several observed events. 

For several TDEs, the slow evolution of the radio light curves favors higher flare masses, often from Flare+$\alpha0$CNM models. Figure \ref{fig:rise_decay_times} demonstrates that for the  Flare+$\alpha0$CNM models, lower flare masses lead to light curves that rise and fall much more quickly than for higher mass flares. Decreasing the flare velocity could increase the rise time, but would also create a dimmer light curve. Thus, simply using the flare mass as indicated by the lower inferred $M_{\rm BH}$ in the \cite{piro2025} framework cannot easily reproduce the bright, slowly evolving radio emission of several events that prefer our models with larger flare masses.

In the models of \cite{piro2025}, the disk mass and outflow timescales do vary with different assumptions for the disk viscosity and mass inflow rate, perhaps allowing for larger flare masses even for lower mass black holes. For instance, shallower power-law inflow rates $s$ can lead to less compact disks and larger flare masses. A smaller disk viscosity allows the disk to evolve more slowly, leading to fewer outbursts but more mass ejected per flare. In addition, the predictions from the one-zone disk model are subject to significant quantitative uncertainty, and future calculations with a more sophisticated model would elucidate whether the flare masses that we use to model certain events are reasonable to expect. 

Nevertheless, the framework used to calculate the dynamics of the flare collisions and their radio emission is mostly agnostic to the exact process that produces these mass outflows of $\sim 0.01$--$0.1\, M_{\odot}$. We demonstrate generally that such ejecta can produce bright radio emission when they collide either with the CNM or with each other. For the specific density profile and velocity range that we have assumed for each outflow, which are motivated by the expected properties of a radiatively inefficient accretion disk around a SMBH, the shocks formed by the collisions power a flux of $L_{\nu} \sim 10^{27}$--$10^{30}\,  \mathrm{erg}\, \mathrm{s}^{-1}\, \mathrm{Hz}^{-1}$ at 6 GHz. Interaction with the CNM can slow the shock to $\lesssim 0.1c$ at radii of $\sim 10^{17}$--$10^{18}$ cm, whereas the collision of two flares produces a shocked shell moving at $\sim 0.1c$ out to large radii. If the outflow were instead ejected with a smaller maximum velocity or had a steeper density profile $s>0.5$, the predicted flux would tend to be dimmer, as both these changes cause a slower shock velocity. A flatter outflow density profile would produce brighter light curves with shorter peak timescales, as more material in the outflow would move at higher velocity. 

Another aspect of the model that may warrant further exploration is the variation of the light curve with a different power law index of the CNM. Other than a constant density CNM, we also chose to explore $\alpha=2.5$ based on, e.g., \cite{alexander2016,cendes2021,goodwin2022}. However, other studies have inferred shallower power laws of $\rho \propto r^{-1}$ \citep{cendes2022,goodwin2023,burn2025arXiv}. Models with a less steep power law index $\alpha$ would occupy the space between the Flare+$\alpha2.5$CNM and Flare+$\alpha0$CNM models in Figure \ref{fig:rise_decay_times}. More late-time radio observations of TDEs will further clarify the expected range of CNM density profiles.

\subsection{Summary and Conclusions}

In this work, we study the radio emission produced by collisions of outflows, either with each other or with dense CNM. Our outflow properties are motivated by the scenario of disk instabilities leading to flaring events and mass ejection when the disk is in a high accretion state \citep[e.g.][]{piro2025}. Overall, our results for the hydrodynamical evolution and the radio emission from the flare collision scenarios are consistent with the range of inferred properties for observed late-time radio emission in TDEs \citep{cendes2024,goodwin2025arXiv}. 

Our analysis demonstrates that collisions of two flares with each other (Flare+Flare models) tend to produce bright radio light curves that steeply rise and decline on $\sim 10$--$10^2$ d timescales, and this scenario has the potential to explain delayed radio emission that exhibits multiple rapidly evolving peaks in the light curve. The collision of flares with dense CNM can decay on much longer timescales ($\sim 10^2$--$10^3$ d) for flat CNM density profiles (Flare+$\alpha0$CNM models), but evolves more rapidly for steeper CNM density profiles (Flare+$\alpha2.5$CNM models). Comparison of our models with specific TDEs yields promising agreement with both light curves and SEDs for many events, lending credence to the idea that delayed, mildly-relativistic outflows underlie the late-time radio emission discovered for many TDEs.

Ultimately, late time radio emission in TDEs may derive from a variety of origins. While we find that the scenario of delayed disk outflows leading to flare collisions constitutes a promising explanation for several events, our work also highlights the benefits of more late-time radio data in order to distinguish between different types of models for generating the delayed radio emission. In particular, resolved light curve peaks and multi-frequency observations that constrain the SEDs would motivate further quantitative comparisons between models and data. 
Future multiwavelength monitoring of TDEs in the years to decades after discovery will have the power to illuminate even more details of the late-time evolution of these fascinating transients.

\begin{acknowledgments}
    We thank the anonymous referee for their valuable feedback, which has significantly improved the paper. We are grateful to Assaf Horesh, Xiaoshan Huang and Itai Sfaradi for helpful discussions. D.T. is supported by the Sherman Fairchild Postdoctoral Fellowship at Caltech.
\end{acknowledgments}


\bibliography{bib}

@ARTICLE{mummery2021,
	title = {Tidal disruption event discs are larger than they seem: removing systematic biases in {TDE} {X}-ray spectral modelling},
	volume = {507},
	issn = {1745-3925},
	shorttitle = {Tidal disruption event discs are larger than they seem},
	url = {https://doi.org/10.1093/mnrasl/slab088},
	doi = {10.1093/mnrasl/slab088},
	number = {1},
	urldate = {2023-04-10},
	journal = {Monthly Notices of the Royal Astronomical Society: Letters},
	author = {Mummery, Andrew},
	month = oct,
	year = {2021},
	pages = {L24--L28},
}

@ARTICLE{goodwin2025arXiv,
       author = {{Goodwin}, A.~J. and {Burn}, M. and {Anderson}, G.~E. and {Miller-Jones}, J.~C.~A. and {Grotova}, I. and {Baldini}, P. and {Liu}, Z. and {Malyali}, A. and {Rau}, A. and {Salvato}, M.},
        title = "{A systematic analysis of the radio properties of 22 X-ray selected tidal disruption event candidates with the Australia Telescope Compact Array}",
      journal = {arXiv e-prints},
         year = 2025,
        month = apr,
          eid = {arXiv:2504.08426},
        pages = {arXiv:2504.08426},
          doi = {10.48550/arXiv.2504.08426},
archivePrefix = {arXiv},
       eprint = {2504.08426},
 primaryClass = {astro-ph.HE},
       adsurl = {https://ui.adsabs.harvard.edu/abs/2025arXiv250408426G}
}

@ARTICLE{piro2025,
       author = {{Piro}, Anthony L. and {Mockler}, Brenna},
        title = "{Late-time Evolution and Instabilities of Tidal Disruption Disks}",
      journal = {\apj},
         year = 2025,
        month = may,
       volume = {985},
       number = {1},
          eid = {77},
        pages = {77},
          doi = {10.3847/1538-4357/adc729},
archivePrefix = {arXiv},
       eprint = {2412.01922},
 primaryClass = {astro-ph.HE},
       adsurl = {https://ui.adsabs.harvard.edu/abs/2025ApJ...985...77P}
}

@ARTICLE{cendes2024,
       author = {{Cendes}, Y. and {Berger}, E. and {Alexander}, K.~D. and {Chornock}, R. and {Margutti}, R. and {Metzger}, B. and {Wieringa}, M.~H. and {Bietenholz}, M.~F. and {Hajela}, A. and {Laskar}, T. and {Stroh}, M.~C. and {Terreran}, G.},
        title = "{Ubiquitous Late Radio Emission from Tidal Disruption Events}",
      journal = {\apj},
         year = 2024,
        month = aug,
       volume = {971},
       number = {2},
          eid = {185},
        pages = {185},
          doi = {10.3847/1538-4357/ad5541},
archivePrefix = {arXiv},
       eprint = {2308.13595},
 primaryClass = {astro-ph.HE},
       adsurl = {https://ui.adsabs.harvard.edu/abs/2024ApJ...971..185C}
}

@ARTICLE{sfaradi2025,
       author = {{Sfaradi}, Itai and {Margutti}, Raffaella and {Chornock}, Ryan and {Alexander}, Kate D. and {Metzger}, Brian D. and {Beniamini}, Paz and {Duran}, Rodolfo Barniol and {Yao}, Yuhan and {Horesh}, Assaf and {Farah}, Wael and {Berger}, Edo and {A.~J.}, Nayana and {Cendes}, Yvette and {Eftekhari}, Tarraneh and {Fender}, Rob and {Franz}, Noah and {Green}, Dave A. and {Hammerstein}, Erica and {Lu}, Wenbin and {Wiston}, Eli and {Bernstein}, Yirmi and {Bright}, Joe and {Christy}, Collin T. and {Cruz}, Luigi F. and {DeBoer}, David R. and {Golay}, Walter W. and {Goodwin}, Adelle J. and {Gurwell}, Mark and {Keating}, Garrett K. and {Laskar}, Tanmoy and {Miller-Jones}, James C.~A. and {Pollak}, Alexander W. and {Rao}, Ramprasad and {Siemion}, Andrew and {Sheikh}, Sofia Z. and {Shoval}, Nadav and {van Velzen}, Sjoert},
        title = "{The First Radio-bright Off-nuclear Tidal Disruption Event AT 2024tvd Reveals the Fastest-evolving Double-peaked Radio Emission}",
      journal = {\apjl},
         year = 2025,
        month = oct,
       volume = {992},
       number = {2},
          eid = {L18},
        pages = {L18},
          doi = {10.3847/2041-8213/ae0a26},
       adsurl = {https://ui.adsabs.harvard.edu/abs/2025ApJ...992L..18S}
}

@ARTICLE{vanvelzen19,
       author = {{van Velzen}, Sjoert and {Stone}, Nicholas C. and {Metzger}, Brian D. and {Gezari}, Suvi and {Brown}, Thomas M. and {Fruchter}, Andrew S.},
        title = "{Late-time UV Observations of Tidal Disruption Flares Reveal Unobscured, Compact Accretion Disks}",
      journal = {\apj},
         year = 2019,
        month = jun,
       volume = {878},
       number = {2},
          eid = {82},
        pages = {82},
          doi = {10.3847/1538-4357/ab1844},
archivePrefix = {arXiv},
       eprint = {1809.00003},
 primaryClass = {astro-ph.HE},
       adsurl = {https://ui.adsabs.harvard.edu/abs/2019ApJ...878...82V}
}

@ARTICLE{hu2025,
       author = {{Hu}, Fangyi (Fitz) and {Goodwin}, Adelle and {Price}, Daniel J. and {Mandel}, Ilya and {Sari}, Re'em and {Hayasaki}, Kimitake},
        title = "{Radio Emission from Tidal Disruption Events Produced by the Collision between Super-Eddington Outflows and the Circumnuclear Medium}",
      journal = {\apjl},
         year = 2025,
        month = jul,
       volume = {988},
       number = {1},
          eid = {L24},
        pages = {L24},
          doi = {10.3847/2041-8213/adeb79},
archivePrefix = {arXiv},
       eprint = {2507.01273},
 primaryClass = {astro-ph.HE},
       adsurl = {https://ui.adsabs.harvard.edu/abs/2025ApJ...988L..24H}
}

@ARTICLE{sfaradi2022,
       author = {{Sfaradi}, Itai and {Horesh}, Assaf and {Fender}, Rob and {Green}, David A. and {Williams}, David R.~A. and {Bright}, Joe and {Schulze}, Steve},
        title = "{A Late-time Radio Flare Following a Possible Transition in Accretion State in the Tidal Disruption Event AT 2019azh}",
      journal = {\apj},
         year = 2022,
        month = jul,
       volume = {933},
       number = {2},
          eid = {176},
        pages = {176},
          doi = {10.3847/1538-4357/ac74bc},
archivePrefix = {arXiv},
       eprint = {2202.00026},
 primaryClass = {astro-ph.HE},
       adsurl = {https://ui.adsabs.harvard.edu/abs/2022ApJ...933..176S}
}

@ARTICLE{Nakar11,
       author = {{Nakar}, Ehud and {Piran}, Tsvi},
        title = "{Detectable radio flares following gravitational waves from mergers of binary neutron stars}",
      journal = {\nat},
         year = 2011,
        month = oct,
       volume = {478},
       number = {7367},
        pages = {82-84},
          doi = {10.1038/nature10365},
archivePrefix = {arXiv},
       eprint = {1102.1020},
 primaryClass = {astro-ph.HE},
       adsurl = {https://ui.adsabs.harvard.edu/abs/2011Natur.478...82N}
}

@ARTICLE{alexander25,
       author = {{Alexander}, Kate D. and {Margutti}, Raffaella and {Gomez}, Sebastian and {Stroh}, Michael and {Chornock}, Ryan and {Laskar}, Tanmoy and {Cendes}, Y. and {Berger}, Edo and {Eftekhari}, Tarraneh and {Franz}, Noah and {Hajela}, Aprajita and {Metzger}, B.~D. and {Terreran}, Giacomo and {Bietenholz}, Michael and {Christy}, Collin and {de Colle}, Fabio and {Komossa}, S. and {Nicholl}, Matt and {Ramirez-Ruiz}, Enrico and {Saxton}, Richard and {Schroeder}, Genevieve and {Williams}, Peter and {Wu}, William},
        title = "{The Multi-Wavelength Context of Delayed Radio Emission in TDEs: Evidence for Accretion-Driven Outflows}",
      journal = {arXiv e-prints},
         year = 2025,
        month = jun,
          eid = {arXiv:2506.12729},
        pages = {arXiv:2506.12729},
          doi = {10.48550/arXiv.2506.12729},
archivePrefix = {arXiv},
       eprint = {2506.12729},
 primaryClass = {astro-ph.HE},
       adsurl = {https://ui.adsabs.harvard.edu/abs/2025arXiv250612729A}
}

@ARTICLE{hajela2025,
       author = {{Hajela}, A. and {Alexander}, K.~D. and {Margutti}, R. and {Chornock}, R. and {Bietenholz}, M. and {Christy}, C.~T. and {Stroh}, M. and {Terreran}, G. and {Saxton}, R. and {Komossa}, S. and {Bright}, J.~S. and {Ramirez-Ruiz}, E. and {Coppejans}, D.~L. and {Leung}, J.~K. and {Cendes}, Y. and {Wiston}, E. and {Laskar}, T. and {Horesh}, A. and {Schroeder}, G. and {A.~J.}, Nayana and {Wieringa}, M.~H. and {Velez}, N. and {Berger}, E. and {Blanchard}, P.~K. and {Eftekhari}, T. and {Gomez}, S. and {Nicholl}, M. and {Sears}, H. and {Zauderer}, B.~A.},
        title = "{Eight Years of Light from ASASSN-15oi: Toward Understanding the Late-time Evolution of TDEs}",
      journal = {\apj},
         year = 2025,
        month = apr,
       volume = {983},
       number = {1},
          eid = {29},
        pages = {29},
          doi = {10.3847/1538-4357/adb620},
archivePrefix = {arXiv},
       eprint = {2407.19019},
 primaryClass = {astro-ph.HE},
       adsurl = {https://ui.adsabs.harvard.edu/abs/2025ApJ...983...29H}
}

@ARTICLE{horesh2021,
       author = {{Horesh}, A. and {Cenko}, S.~B. and {Arcavi}, I.},
        title = "{Delayed radio flares from a tidal disruption event}",
      journal = {Nature Astronomy},
         year = 2021,
        month = may,
       volume = {5},
        pages = {491-497},
          doi = {10.1038/s41550-021-01300-8},
archivePrefix = {arXiv},
       eprint = {2102.11290},
 primaryClass = {astro-ph.HE},
       adsurl = {https://ui.adsabs.harvard.edu/abs/2021NatAs...5..491H}
}

@ARTICLE{lu2020,
       author = {{Lu}, Wenbin and {Bonnerot}, Cl{\'e}ment},
        title = "{Self-intersection of the fallback stream in tidal disruption events}",
      journal = {\mnras},
         year = 2020,
        month = feb,
       volume = {492},
       number = {1},
        pages = {686-707},
          doi = {10.1093/mnras/stz3405},
archivePrefix = {arXiv},
       eprint = {1904.12018},
 primaryClass = {astro-ph.HE},
       adsurl = {https://ui.adsabs.harvard.edu/abs/2020MNRAS.492..686L}
}

@ARTICLE{yao2025,
       author = {{Yao}, Yuhan and {Chornock}, Ryan and {Ward}, Charlotte and {Hammerstein}, Erica and {Sfaradi}, Itai and {Margutti}, Raffaella and {Kelley}, Luke Zoltan and {Lu}, Wenbin and {Liu}, Chang and {Wise}, Jacob and {Sollerman}, Jesper and {Alexander}, Kate D. and {Bellm}, Eric C. and {Drake}, Andrew J. and {Fremling}, Christoffer and {Gilfanov}, Marat and {Graham}, Matthew J. and {Groom}, Steven L. and {Hinds}, K.~R. and {Kulkarni}, S.~R. and {Miller}, Adam A. and {Miller-Jones}, James C.~A. and {Nicholl}, Matt and {Perley}, Daniel A. and {Purdum}, Josiah and {Ravi}, Vikram and {Rich}, R. Michael and {Rehemtulla}, Nabeel and {Riddle}, Reed and {Smith}, Roger and {Stein}, Robert and {Sunyaev}, Rashid and {van Velzen}, Sjoert and {Wold}, Avery},
        title = "{A Massive Black Hole 0.8 kpc from the Host Nucleus Revealed by the Offset Tidal Disruption Event AT2024tvd}",
      journal = {\apjl},
         year = 2025,
        month = jun,
       volume = {985},
       number = {2},
          eid = {L48},
        pages = {L48},
          doi = {10.3847/2041-8213/add7de},
archivePrefix = {arXiv},
       eprint = {2502.17661},
 primaryClass = {astro-ph.GA},
       adsurl = {https://ui.adsabs.harvard.edu/abs/2025ApJ...985L..48Y}
}

@ARTICLE{cendes2021,
       author = {{Cendes}, Y. and {Alexander}, K.~D. and {Berger}, E. and {Eftekhari}, T. and {Williams}, P.~K.~G. and {Chornock}, R.},
        title = "{Radio Observations of an Ordinary Outflow from the Tidal Disruption Event AT2019dsg}",
      journal = {\apj},
         year = 2021,
        month = oct,
       volume = {919},
       number = {2},
          eid = {127},
        pages = {127},
          doi = {10.3847/1538-4357/ac110a},
archivePrefix = {arXiv},
       eprint = {2103.06299},
 primaryClass = {astro-ph.HE},
       adsurl = {https://ui.adsabs.harvard.edu/abs/2021ApJ...919..127C}
}

@ARTICLE{alexander2016,
       author = {{Alexander}, K.~D. and {Berger}, E. and {Guillochon}, J. and {Zauderer}, B.~A. and {Williams}, P.~K.~G.},
        title = "{Discovery of an Outflow from Radio Observations of the Tidal Disruption Event ASASSN-14li}",
      journal = {\apjl},
         year = 2016,
        month = mar,
       volume = {819},
       number = {2},
          eid = {L25},
        pages = {L25},
          doi = {10.3847/2041-8205/819/2/L25},
archivePrefix = {arXiv},
       eprint = {1510.01226},
 primaryClass = {astro-ph.HE},
       adsurl = {https://ui.adsabs.harvard.edu/abs/2016ApJ...819L..25A}
}

@ARTICLE{Chevalier98,
       author = {{Chevalier}, Roger A.},
        title = "{Synchrotron Self-Absorption in Radio Supernovae}",
      journal = {\apj},
         year = 1998,
        month = may,
       volume = {499},
       number = {2},
        pages = {810-819},
          doi = {10.1086/305676},
       adsurl = {https://ui.adsabs.harvard.edu/abs/1998ApJ...499..810C}
}

@ARTICLE{guo2024,
       author = {{Guo}, Minghao and {Stone}, James M. and {Quataert}, Eliot and {Kim}, Chang-Goo},
        title = "{Magnetized Accretion onto and Feedback from Supermassive Black Holes in Elliptical Galaxies}",
      journal = {\apj},
         year = 2024,
        month = oct,
       volume = {973},
       number = {2},
          eid = {141},
        pages = {141},
          doi = {10.3847/1538-4357/ad5fe7},
archivePrefix = {arXiv},
       eprint = {2405.11711},
 primaryClass = {astro-ph.HE},
       adsurl = {https://ui.adsabs.harvard.edu/abs/2024ApJ...973..141G}
}

@ARTICLE{cho2024,
       author = {{Cho}, Hyerin and {Prather}, Ben S. and {Su}, Kung-Yi and {Narayan}, Ramesh and {Natarajan}, Priyamvada},
        title = "{Multizone Modeling of Black Hole Accretion and Feedback in 3D GRMHD: Bridging Vast Spatial and Temporal Scales}",
      journal = {\apj},
         year = 2024,
        month = dec,
       volume = {977},
       number = {2},
          eid = {200},
        pages = {200},
          doi = {10.3847/1538-4357/ad9561},
archivePrefix = {arXiv},
       eprint = {2405.13887},
 primaryClass = {astro-ph.HE},
       adsurl = {https://ui.adsabs.harvard.edu/abs/2024ApJ...977..200C}
}

@ARTICLE{burn2025arXiv,
       author = {{Burn}, Matthew and {Goodwin}, Adelle J. and {Anderson}, Gemma E. and {Miller-Jones}, James C.~A. and {Cendes}, Yvette and {Christy}, Collin T. and {Lu}, Wenbin and {van Velzen}, Sjoert},
        title = "{The 6 year radio lightcurve of the tidal disruption event AT2019azh}",
      journal = {arXiv e-prints},
         year = 2025,
        month = sep,
          eid = {arXiv:2509.17525},
        pages = {arXiv:2509.17525},
archivePrefix = {arXiv},
       eprint = {2509.17525},
 primaryClass = {astro-ph.HE},
       adsurl = {https://ui.adsabs.harvard.edu/abs/2025arXiv250917525B}
}

@ARTICLE{goodwin2025,
       author = {{Goodwin}, A.~J. and {Mummery}, A. and {Laskar}, T. and {Alexander}, K.~D. and {Anderson}, G.~E. and {Bietenholz}, M. and {Bonnerot}, C. and {Christy}, C.~T. and {Golay}, W. and {Lu}, W. and {Margutti}, R. and {Miller-Jones}, J.~C.~A. and {Ramirez-Ruiz}, E. and {Saxton}, R. and {van Velzen}, S.},
        title = "{A Second Radio Flare from the Tidal Disruption Event AT2020vwl: A Delayed Outflow Ejection?}",
      journal = {\apj},
         year = 2025,
        month = mar,
       volume = {981},
       number = {2},
          eid = {122},
        pages = {122},
          doi = {10.3847/1538-4357/adb0b1},
archivePrefix = {arXiv},
       eprint = {2410.18665},
 primaryClass = {astro-ph.HE},
       adsurl = {https://ui.adsabs.harvard.edu/abs/2025ApJ...981..122G}
}

@ARTICLE{goodwin2023,
       author = {{Goodwin}, A.~J. and {Alexander}, K.~D. and {Miller-Jones}, J.~C.~A. and {Bietenholz}, M.~F. and {van Velzen}, S. and {Anderson}, G.~E. and {Berger}, E. and {Cendes}, Y. and {Chornock}, R. and {Coppejans}, D.~L. and {Eftekhari}, T. and {Gezari}, S. and {Laskar}, T. and {Ramirez-Ruiz}, E. and {Saxton}, R.},
        title = "{A radio-emitting outflow produced by the tidal disruption event AT2020vwl}",
      journal = {\mnras},
         year = 2023,
        month = jul,
       volume = {522},
       number = {4},
        pages = {5084-5097},
          doi = {10.1093/mnras/stad1258},
archivePrefix = {arXiv},
       eprint = {2304.12661},
 primaryClass = {astro-ph.HE},
       adsurl = {https://ui.adsabs.harvard.edu/abs/2023MNRAS.522.5084G}
}

@ARTICLE{wevers2021,
       author = {{Wevers}, T. and {Pasham}, D.~R. and {van Velzen}, S. and {Miller-Jones}, J.~C.~A. and {Uttley}, P. and {Gendreau}, K.~C. and {Remillard}, R. and {Arzoumanian}, Z. and {L{\"o}wenstein}, M. and {Chiti}, A.},
        title = "{Rapid Accretion State Transitions following the Tidal Disruption Event AT2018fyk}",
      journal = {\apj},
         year = 2021,
        month = may,
       volume = {912},
       number = {2},
          eid = {151},
        pages = {151},
          doi = {10.3847/1538-4357/abf5e2},
archivePrefix = {arXiv},
       eprint = {2101.04692},
 primaryClass = {astro-ph.HE},
       adsurl = {https://ui.adsabs.harvard.edu/abs/2021ApJ...912..151W}
}

@ARTICLE{yang2025,
       author = {{Yang}, Lei and {Shu}, Xinwen and {Mou}, Goubin and {Xue}, Yongquan and {Sun}, Luming and {Zhang}, Fabao and {Zhang}, Zhumao and {Wang}, Yibo and {Wu}, Tao and {Jiang}, Ning and {Ding}, Hucheng and {Wang}, Tinggui},
        title = "{Outflow-cloud interaction as the possible origin of the peculiar radio emission in the tidal disruption event AT2018cqh}",
      journal = {arXiv e-prints},
         year = 2025,
        month = sep,
          eid = {arXiv:2509.21299},
        pages = {arXiv:2509.21299},
          doi = {10.48550/arXiv.2509.21299},
archivePrefix = {arXiv},
       eprint = {2509.21299},
 primaryClass = {astro-ph.HE},
       adsurl = {https://ui.adsabs.harvard.edu/abs/2025arXiv250921299Y}
}

@ARTICLE{cendes2022,
       author = {{Cendes}, Y. and {Berger}, E. and {Alexander}, K.~D. and {Gomez}, S. and {Hajela}, A. and {Chornock}, R. and {Laskar}, T. and {Margutti}, R. and {Metzger}, B. and {Bietenholz}, M.~F. and {Brethauer}, D. and {Wieringa}, M.~H.},
        title = "{A Mildly Relativistic Outflow Launched Two Years after Disruption in Tidal Disruption Event AT2018hyz}",
      journal = {\apj},
         year = 2022,
        month = oct,
       volume = {938},
       number = {1},
          eid = {28},
        pages = {28},
          doi = {10.3847/1538-4357/ac88d0},
archivePrefix = {arXiv},
       eprint = {2206.14297},
 primaryClass = {astro-ph.HE},
       adsurl = {https://ui.adsabs.harvard.edu/abs/2022ApJ...938...28C}
}

@ARTICLE{rees1988,
       author = {{Rees}, Martin J.},
        title = "{Tidal disruption of stars by black holes of {}10$^{6}$-{}10$^{8}$ solar masses in nearby galaxies}",
      journal = {\nat},
         year = 1988,
        month = jun,
       volume = {333},
       number = {6173},
        pages = {523-528},
          doi = {10.1038/333523a0},
       adsurl = {https://ui.adsabs.harvard.edu/abs/1988Natur.333..523R}
}

@ARTICLE{hills1975,
       author = {{Hills}, J.~G.},
        title = "{Possible power source of Seyfert galaxies and QSOs}",
      journal = {\nat},
         year = 1975,
        month = mar,
       volume = {254},
       number = {5498},
        pages = {295-298},
          doi = {10.1038/254295a0},
       adsurl = {https://ui.adsabs.harvard.edu/abs/1975Natur.254..295H}
}

@ARTICLE{evans1989,
       author = {{Evans}, Charles R. and {Kochanek}, Christopher S.},
        title = "{The Tidal Disruption of a Star by a Massive Black Hole}",
      journal = {\apjl},
         year = 1989,
        month = nov,
       volume = {346},
        pages = {L13},
          doi = {10.1086/185567},
       adsurl = {https://ui.adsabs.harvard.edu/abs/1989ApJ...346L..13E}
}

@ARTICLE{guillochon2013,
       author = {{Guillochon}, James and {Ramirez-Ruiz}, Enrico},
        title = "{Hydrodynamical Simulations to Determine the Feeding Rate of Black Holes by the Tidal Disruption of Stars: The Importance of the Impact Parameter and Stellar Structure}",
      journal = {\apj},
         year = 2013,
        month = apr,
       volume = {767},
       number = {1},
          eid = {25},
        pages = {25},
          doi = {10.1088/0004-637X/767/1/25},
archivePrefix = {arXiv},
       eprint = {1206.2350},
 primaryClass = {astro-ph.HE},
       adsurl = {https://ui.adsabs.harvard.edu/abs/2013ApJ...767...25G}
}

@ARTICLE{alexander2020,
       author = {{Alexander}, Kate D. and {van Velzen}, Sjoert and {Horesh}, Assaf and {Zauderer}, B. Ashley},
        title = "{Radio Properties of Tidal Disruption Events}",
      journal = {\ssr},
         year = 2020,
        month = jun,
       volume = {216},
       number = {5},
          eid = {81},
        pages = {81},
          doi = {10.1007/s11214-020-00702-w},
archivePrefix = {arXiv},
       eprint = {2006.01159},
 primaryClass = {astro-ph.HE},
       adsurl = {https://ui.adsabs.harvard.edu/abs/2020SSRv..216...81A}
}

@ARTICLE{zauderer2011,
       author = {{Zauderer}, B.~A. and {Berger}, E. and {Soderberg}, A.~M. and {Loeb}, A. and {Narayan}, R. and {Frail}, D.~A. and {Petitpas}, G.~R. and {Brunthaler}, A. and {Chornock}, R. and {Carpenter}, J.~M. and {Pooley}, G.~G. and {Mooley}, K. and {Kulkarni}, S.~R. and {Margutti}, R. and {Fox}, D.~B. and {Nakar}, E. and {Patel}, N.~A. and {Volgenau}, N.~H. and {Culverhouse}, T.~L. and {Bietenholz}, M.~F. and {Rupen}, M.~P. and {Max-Moerbeck}, W. and {Readhead}, A.~C.~S. and {Richards}, J. and {Shepherd}, M. and {Storm}, S. and {Hull}, C.~L.~H.},
        title = "{Birth of a relativistic outflow in the unusual {\ensuremath{\gamma}}-ray transient Swift J164449.3+573451}",
      journal = {\nat},
         year = 2011,
        month = aug,
       volume = {476},
       number = {7361},
        pages = {425-428},
          doi = {10.1038/nature10366},
archivePrefix = {arXiv},
       eprint = {1106.3568},
 primaryClass = {astro-ph.HE},
       adsurl = {https://ui.adsabs.harvard.edu/abs/2011Natur.476..425Z}
}

@ARTICLE{zauderer2013,
       author = {{Zauderer}, B.~A. and {Berger}, E. and {Margutti}, R. and {Pooley}, G.~G. and {Sari}, R. and {Soderberg}, A.~M. and {Brunthaler}, A. and {Bietenholz}, M.~F.},
        title = "{Radio Monitoring of the Tidal Disruption Event Swift J164449.3+573451. II. The Relativistic Jet Shuts Off and a Transition to Forward Shock X-Ray/Radio Emission}",
      journal = {\apj},
         year = 2013,
        month = apr,
       volume = {767},
       number = {2},
          eid = {152},
        pages = {152},
          doi = {10.1088/0004-637X/767/2/152},
archivePrefix = {arXiv},
       eprint = {1212.1173},
 primaryClass = {astro-ph.HE},
       adsurl = {https://ui.adsabs.harvard.edu/abs/2013ApJ...767..152Z}
}

@ARTICLE{matsumoto2023,
       author = {{Matsumoto}, Tatsuya and {Piran}, Tsvi},
        title = "{Generalized equipartition method from an arbitrary viewing angle}",
      journal = {\mnras},
         year = 2023,
        month = jul,
       volume = {522},
       number = {3},
        pages = {4565-4576},
          doi = {10.1093/mnras/stad1269},
archivePrefix = {arXiv},
       eprint = {2211.10051},
 primaryClass = {astro-ph.HE},
       adsurl = {https://ui.adsabs.harvard.edu/abs/2023MNRAS.522.4565M}
}

@ARTICLE{sfaradi2024,
       author = {{Sfaradi}, Itai and {Beniamini}, Paz and {Horesh}, Assaf and {Piran}, Tsvi and {Bright}, Joe and {Rhodes}, Lauren and {Williams}, David R.~A. and {Fender}, Rob and {Leung}, James K. and {Murphy}, Tara and {Green}, Dave A.},
        title = "{An off-axis relativistic jet seen in the long lasting delayed radio flare of the TDE AT 2018hyz}",
      journal = {\mnras},
         year = 2024,
        month = jan,
       volume = {527},
       number = {3},
        pages = {7672-7680},
          doi = {10.1093/mnras/stad3717},
archivePrefix = {arXiv},
       eprint = {2308.01965},
 primaryClass = {astro-ph.HE},
       adsurl = {https://ui.adsabs.harvard.edu/abs/2024MNRAS.527.7672S}
}

@ARTICLE{teboul2023,
       author = {{Teboul}, Odelia and {Metzger}, Brian D.},
        title = "{A Unified Theory of Jetted Tidal Disruption Events: From Promptly Escaping Relativistic to Delayed Transrelativistic Jets}",
      journal = {\apjl},
         year = 2023,
        month = nov,
       volume = {957},
       number = {1},
          eid = {L9},
        pages = {L9},
          doi = {10.3847/2041-8213/ad0037},
archivePrefix = {arXiv},
       eprint = {2308.05161},
 primaryClass = {astro-ph.HE},
       adsurl = {https://ui.adsabs.harvard.edu/abs/2023ApJ...957L...9T}
}

@ARTICLE{lu2024,
       author = {{Lu}, Wenbin and {Matsumoto}, Tatsuya and {Matzner}, Christopher D.},
        title = "{Misaligned precessing jets are choked by the accretion disc wind}",
      journal = {\mnras},
         year = 2024,
        month = sep,
       volume = {533},
       number = {1},
        pages = {979-993},
          doi = {10.1093/mnras/stae1770},
archivePrefix = {arXiv},
       eprint = {2310.15336},
 primaryClass = {astro-ph.HE},
       adsurl = {https://ui.adsabs.harvard.edu/abs/2024MNRAS.533..979L}
}

@ARTICLE{mockler2019,
       author = {{Mockler}, Brenna and {Guillochon}, James and {Ramirez-Ruiz}, Enrico},
        title = "{Weighing Black Holes Using Tidal Disruption Events}",
      journal = {\apj},
         year = 2019,
        month = feb,
       volume = {872},
       number = {2},
          eid = {151},
        pages = {151},
          doi = {10.3847/1538-4357/ab010f},
archivePrefix = {arXiv},
       eprint = {1801.08221},
 primaryClass = {astro-ph.HE},
       adsurl = {https://ui.adsabs.harvard.edu/abs/2019ApJ...872..151M}
}

@ARTICLE{auchettl2017,
       author = {{Auchettl}, Katie and {Guillochon}, James and {Ramirez-Ruiz}, Enrico},
        title = "{New Physical Insights about Tidal Disruption Events from a Comprehensive Observational Inventory at X-Ray Wavelengths}",
      journal = {\apj},
         year = 2017,
        month = apr,
       volume = {838},
       number = {2},
          eid = {149},
        pages = {149},
          doi = {10.3847/1538-4357/aa633b},
archivePrefix = {arXiv},
       eprint = {1611.02291},
 primaryClass = {astro-ph.HE},
       adsurl = {https://ui.adsabs.harvard.edu/abs/2017ApJ...838..149A}
}

@ARTICLE{andreoni2022,
       author = {{Andreoni}, Igor and {Coughlin}, Michael W. and {Perley}, Daniel A. and {Yao}, Yuhan and {Lu}, Wenbin and {Cenko}, S. Bradley and {Kumar}, Harsh and {Anand}, Shreya and {Ho}, Anna Y.~Q. and {Kasliwal}, Mansi M. and {de Ugarte Postigo}, Antonio and {Sagu{\'e}s-Carracedo}, Ana and {Schulze}, Steve and {Kann}, D. Alexander and {Kulkarni}, S.~R. and {Sollerman}, Jesper and {Tanvir}, Nial and {Rest}, Armin and {Izzo}, Luca and {Somalwar}, Jean J. and {Kaplan}, David L. and {Ahumada}, Tom{\'a}s and {Anupama}, G.~C. and {Auchettl}, Katie and {Barway}, Sudhanshu and {Bellm}, Eric C. and {Bhalerao}, Varun and {Bloom}, Joshua S. and {Bremer}, Michael and {Bulla}, Mattia and {Burns}, Eric and {Campana}, Sergio and {Chandra}, Poonam and {Charalampopoulos}, Panos and {Cooke}, Jeff and {D'Elia}, Valerio and {Das}, Kaustav Kashyap and {Dobie}, Dougal and {Ag{\"u}{\'\i} Fern{\'a}ndez}, Jos{\'e} Feliciano and {Freeburn}, James and {Fremling}, Cristoffer and {Gezari}, Suvi and {Goode}, Simon and {Graham}, Matthew J. and {Hammerstein}, Erica and {Karambelkar}, Viraj R. and {Kilpatrick}, Charles D. and {Kool}, Erik C. and {Krips}, Melanie and {Laher}, Russ R. and {Leloudas}, Giorgos and {Levan}, Andrew and {Lundquist}, Michael J. and {Mahabal}, Ashish A. and {Medford}, Michael S. and {Miller}, M. Coleman and {M{\"o}ller}, Anais and {Mooley}, Kunal P. and {Nayana}, A.~J. and {Nir}, Guy and {Pang}, Peter T.~H. and {Paraskeva}, Emmy and {Perley}, Richard A. and {Petitpas}, Glen and {Pursiainen}, Miika and {Ravi}, Vikram and {Ridden-Harper}, Ryan and {Riddle}, Reed and {Rigault}, Mickael and {Rodriguez}, Antonio C. and {Rusholme}, Ben and {Sharma}, Yashvi and {Smith}, I.~A. and {Stein}, Robert D. and {Th{\"o}ne}, Christina and {Tohuvavohu}, Aaron and {Valdes}, Frank and {van Roestel}, Jan and {Vergani}, Susanna D. and {Wang}, Qinan and {Zhang}, Jielai},
        title = "{A very luminous jet from the disruption of a star by a massive black hole}",
      journal = {\nat},
         year = 2022,
        month = dec,
       volume = {612},
       number = {7940},
        pages = {430-434},
          doi = {10.1038/s41586-022-05465-8},
archivePrefix = {arXiv},
       eprint = {2211.16530},
 primaryClass = {astro-ph.HE},
       adsurl = {https://ui.adsabs.harvard.edu/abs/2022Natur.612..430A}
}

@ARTICLE{hammerstein2023,
       author = {{Hammerstein}, Erica and {van Velzen}, Sjoert and {Gezari}, Suvi and {Cenko}, S. Bradley and {Yao}, Yuhan and {Ward}, Charlotte and {Frederick}, Sara and {Villanueva}, Natalia and {Somalwar}, Jean J. and {Graham}, Matthew J. and {Kulkarni}, Shrinivas R. and {Stern}, Daniel and {Andreoni}, Igor and {Bellm}, Eric C. and {Dekany}, Richard and {Dhawan}, Suhail and {Drake}, Andrew J. and {Fremling}, Christoffer and {Gatkine}, Pradip and {Groom}, Steven L. and {Ho}, Anna Y.~Q. and {Kasliwal}, Mansi M. and {Karambelkar}, Viraj and {Kool}, Erik C. and {Masci}, Frank J. and {Medford}, Michael S. and {Perley}, Daniel A. and {Purdum}, Josiah and {van Roestel}, Jan and {Sharma}, Yashvi and {Sollerman}, Jesper and {Taggart}, Kirsty and {Yan}, Lin},
        title = "{The Final Season Reimagined: 30 Tidal Disruption Events from the ZTF-I Survey}",
      journal = {\apj},
         year = 2023,
        month = jan,
       volume = {942},
       number = {1},
          eid = {9},
        pages = {9},
          doi = {10.3847/1538-4357/aca283},
archivePrefix = {arXiv},
       eprint = {2203.01461},
 primaryClass = {astro-ph.HE},
       adsurl = {https://ui.adsabs.harvard.edu/abs/2023ApJ...942....9H}
}

@ARTICLE{bonnerot2021,
       author = {{Bonnerot}, C. and {Stone}, N.~C.},
        title = "{Formation of an Accretion Flow}",
      journal = {\ssr},
         year = 2021,
        month = feb,
       volume = {217},
       number = {1},
          eid = {16},
        pages = {16},
          doi = {10.1007/s11214-020-00789-1},
archivePrefix = {arXiv},
       eprint = {2008.11731},
 primaryClass = {astro-ph.HE},
       adsurl = {https://ui.adsabs.harvard.edu/abs/2021SSRv..217...16B}
}

@ARTICLE{gezari2021,
       author = {{Gezari}, Suvi},
        title = "{Tidal Disruption Events}",
      journal = {\araa},
         year = 2021,
        month = sep,
       volume = {59},
        pages = {21-58},
          doi = {10.1146/annurev-astro-111720-030029},
archivePrefix = {arXiv},
       eprint = {2104.14580},
 primaryClass = {astro-ph.HE},
       adsurl = {https://ui.adsabs.harvard.edu/abs/2021ARA&A..59...21G}
}

@ARTICLE{zhuang2025,
       author = {{Zhuang}, Jialun and {Shen}, Rong-Feng and {Mou}, Guobin and {Lu}, Wenbin},
        title = "{Interaction of an Outflow with Surrounding Gaseous Clouds as the Origin of Late-time Radio Flares in Tidal Disruption Events}",
      journal = {\apj},
         year = 2025,
        month = feb,
       volume = {979},
       number = {2},
          eid = {109},
        pages = {109},
          doi = {10.3847/1538-4357/ad9b98},
archivePrefix = {arXiv},
       eprint = {2406.08012},
 primaryClass = {astro-ph.HE},
       adsurl = {https://ui.adsabs.harvard.edu/abs/2025ApJ...979..109Z}
}

@ARTICLE{goodwin2022,
       author = {{Goodwin}, A.~J. and {van Velzen}, S. and {Miller-Jones}, J.~C.~A. and {Mummery}, A. and {Bietenholz}, M.~F. and {Wederfoort}, A. and {Hammerstein}, E. and {Bonnerot}, C. and {Hoffmann}, J. and {Yan}, L.},
        title = "{AT2019azh: an unusually long-lived, radio-bright thermal tidal disruption event}",
      journal = {\mnras},
         year = 2022,
        month = apr,
       volume = {511},
       number = {4},
        pages = {5328-5345},
          doi = {10.1093/mnras/stac333},
archivePrefix = {arXiv},
       eprint = {2201.03744},
 primaryClass = {astro-ph.HE},
       adsurl = {https://ui.adsabs.harvard.edu/abs/2022MNRAS.511.5328G}
}

@ARTICLE{blandford99,
       author = {{Blandford}, Roger D. and {Begelman}, Mitchell C.},
        title = "{On the fate of gas accreting at a low rate on to a black hole}",
      journal = {\mnras},
         year = 1999,
        month = feb,
       volume = {303},
       number = {1},
        pages = {L1-L5},
          doi = {10.1046/j.1365-8711.1999.02358.x},
archivePrefix = {arXiv},
       eprint = {astro-ph/9809083},
 primaryClass = {astro-ph},
       adsurl = {https://ui.adsabs.harvard.edu/abs/1999MNRAS.303L...1B}
}

@ARTICLE{greene2020,
       author = {{Greene}, Jenny E. and {Strader}, Jay and {Ho}, Luis C.},
        title = "{Intermediate-Mass Black Holes}",
      journal = {\araa},
         year = 2020,
        month = aug,
       volume = {58},
        pages = {257-312},
          doi = {10.1146/annurev-astro-032620-021835},
archivePrefix = {arXiv},
       eprint = {1911.09678},
 primaryClass = {astro-ph.GA},
       adsurl = {https://ui.adsabs.harvard.edu/abs/2020ARA&A..58..257G}
}

@ARTICLE{wu2025,
       author = {{Wu}, Samantha C. and {Tsuna}, Daichi},
        title = "{Luminous Late-time Radio Emission from Supernovae Interacting with Circumbinary Material}",
      journal = {\apj},
         year = 2025,
        month = dec,
       volume = {994},
       number = {2},
          eid = {141},
        pages = {141},
          doi = {10.3847/1538-4357/ae113c},
archivePrefix = {arXiv},
       eprint = {2507.19613},
 primaryClass = {astro-ph.HE},
       adsurl = {https://ui.adsabs.harvard.edu/abs/2025ApJ...994..141W}
}

@INPROCEEDINGS{phinney1989,
       author = {{Phinney}, E.~S.},
        title = "{Manifestations of a Massive Black Hole in the Galactic Center}",
    booktitle = {The Center of the Galaxy},
         year = 1989,
       editor = {{Morris}, Mark},
       series = {IAU Symposium},
       volume = {136},
        month = jan,
        pages = {543},
       adsurl = {https://ui.adsabs.harvard.edu/abs/1989IAUS..136..543P}
}

@INCOLLECTION{lu2022,
       author = {{Lu}, Wenbin},
        title = "{Accretion Disk Evolution in Tidal Disruption Events}",
    booktitle = {Handbook of X-ray and Gamma-ray Astrophysics},
         year = 2022,
       editor = {{Bambi}, Cosimo and {Sangangelo}, Andrea},
    publisher = {Springer},
          eid = {3},
        pages = {3},
          doi = {10.1007/978-981-16-4544-0_127-1},
       adsurl = {https://ui.adsabs.harvard.edu/abs/2022hxga.book....3L}
}

@ARTICLE{horesh2021b,
       author = {{Horesh}, Assaf and {Sfaradi}, Itai and {Fender}, Rob and {Green}, David A. and {Williams}, David R.~A. and {Bright}, Joe S.},
        title = "{Are Delayed Radio Flares Common in Tidal Disruption Events? The Case of the TDE iPTF 16fnl}",
      journal = {\apjl},
         year = 2021,
        month = oct,
       volume = {920},
       number = {1},
          eid = {L5},
        pages = {L5},
          doi = {10.3847/2041-8213/ac25fe},
archivePrefix = {arXiv},
       eprint = {2109.10921},
 primaryClass = {astro-ph.HE},
       adsurl = {https://ui.adsabs.harvard.edu/abs/2021ApJ...920L...5H}
}

@ARTICLE{shvartzvald2024,
       author = {{Shvartzvald}, Y. and {Waxman}, E. and {Gal-Yam}, A. and {Ofek}, E.~O. and {Ben-Ami}, S. and {Berge}, D. and {Kowalski}, M. and {B{\"u}hler}, R. and {Worm}, S. and {Rhoads}, J.~E. and {Arcavi}, I. and {Maoz}, D. and {Polishook}, D. and {Stone}, N. and {Trakhtenbrot}, B. and {Ackermann}, M. and {Aharonson}, O. and {Birnholtz}, O. and {Chelouche}, D. and {Guetta}, D. and {Hallakoun}, N. and {Horesh}, A. and {Kushnir}, D. and {Mazeh}, T. and {Nordin}, J. and {Ofir}, A. and {Ohm}, S. and {Parsons}, D. and {Pe'er}, A. and {Perets}, H.~B. and {Perdelwitz}, V. and {Poznanski}, D. and {Sadeh}, I. and {Sagiv}, I. and {Shahaf}, S. and {Soumagnac}, M. and {Tal-Or}, L. and {Santen}, J. Van and {Zackay}, B. and {Guttman}, O. and {Rekhi}, P. and {Townsend}, A. and {Weinstein}, A. and {Wold}, I.},
        title = "{ULTRASAT: A Wide-field Time-domain UV Space Telescope}",
      journal = {\apj},
         year = 2024,
        month = mar,
       volume = {964},
       number = {1},
          eid = {74},
        pages = {74},
          doi = {10.3847/1538-4357/ad2704},
archivePrefix = {arXiv},
       eprint = {2304.14482},
 primaryClass = {astro-ph.IM},
       adsurl = {https://ui.adsabs.harvard.edu/abs/2024ApJ...964...74S}
}

@ARTICLE{bonnerot2021b,
       author = {{Bonnerot}, Cl{\'e}ment and {Lu}, Wenbin and {Hopkins}, Philip F.},
        title = "{First light from tidal disruption events}",
      journal = {\mnras},
         year = 2021,
        month = jul,
       volume = {504},
       number = {4},
        pages = {4885-4905},
          doi = {10.1093/mnras/stab398},
archivePrefix = {arXiv},
       eprint = {2012.12271},
 primaryClass = {astro-ph.HE},
       adsurl = {https://ui.adsabs.harvard.edu/abs/2021MNRAS.504.4885B}
}

@ARTICLE{huang2024,
       author = {{Huang}, Xiaoshan and {Davis}, Shane W. and {Jiang}, Yan-fei},
        title = "{Pre-peak Emission in Tidal Disruption Events}",
      journal = {\apj},
         year = 2024,
        month = oct,
       volume = {974},
       number = {2},
          eid = {165},
        pages = {165},
          doi = {10.3847/1538-4357/ad6c39},
archivePrefix = {arXiv},
       eprint = {2404.18446},
 primaryClass = {astro-ph.HE},
       adsurl = {https://ui.adsabs.harvard.edu/abs/2024ApJ...974..165H}
}

@ARTICLE{vanvelzen2016,
       author = {{van Velzen}, S. and {Anderson}, G.~E. and {Stone}, N.~C. and {Fraser}, M. and {Wevers}, T. and {Metzger}, B.~D. and {Jonker}, P.~G. and {van der Horst}, A.~J. and {Staley}, T.~D. and {Mendez}, A.~J. and {Miller-Jones}, J.~C.~A. and {Hodgkin}, S.~T. and {Campbell}, H.~C. and {Fender}, R.~P.},
        title = "{A radio jet from the optical and x-ray bright stellar tidal disruption flare ASASSN-14li}",
      journal = {Science},
         year = 2016,
        month = jan,
       volume = {351},
       number = {6268},
        pages = {62-65},
          doi = {10.1126/science.aad1182},
archivePrefix = {arXiv},
       eprint = {1511.08803},
 primaryClass = {astro-ph.HE},
       adsurl = {https://ui.adsabs.harvard.edu/abs/2016Sci...351...62V}
}

@ARTICLE{christy2024,
       author = {{Christy}, Collin T. and {Alexander}, Kate D. and {Margutti}, Raffaella and {Wieringa}, Mark and {Cendes}, Yvette and {Chornock}, Ryan and {Laskar}, Tanmoy and {Berger}, Edo and {Bietenholz}, Michael and {Coppejans}, Deanne L. and {De Colle}, Fabio and {Eftekhari}, Tarraneh and {Holoien}, Thomas W.-S. and {Matsumoto}, Tatsuya and {Miller-Jones}, James C.~A. and {Ramirez-Ruiz}, Enrico and {Saxton}, Richard and {van Velzen}, Sjoert},
        title = "{The Peculiar Radio Evolution of the Tidal Disruption Event ASASSN-19bt}",
      journal = {\apj},
         year = 2024,
        month = oct,
       volume = {974},
       number = {1},
          eid = {18},
        pages = {18},
          doi = {10.3847/1538-4357/ad675b},
archivePrefix = {arXiv},
       eprint = {2404.12431},
 primaryClass = {astro-ph.HE},
       adsurl = {https://ui.adsabs.harvard.edu/abs/2024ApJ...974...18C}
}

@ARTICLE{cendes2021b,
       author = {{Cendes}, Y. and {Eftekhari}, T. and {Berger}, E. and {Polisensky}, E.},
        title = "{Radio Monitoring of the Tidal Disruption Event Swift J164449.3+573451. IV. Continued Fading and Non-relativistic Expansion}",
      journal = {\apj},
         year = 2021,
        month = feb,
       volume = {908},
       number = {2},
          eid = {125},
        pages = {125},
          doi = {10.3847/1538-4357/abd323},
archivePrefix = {arXiv},
       eprint = {2011.00074},
 primaryClass = {astro-ph.HE},
       adsurl = {https://ui.adsabs.harvard.edu/abs/2021ApJ...908..125C}
}
{}
\bibliographystyle{aasjournalv7}

\end{document}